\documentclass[fleqn,usenatbib]{mnras}
\usepackage{newtxtext,newtxmath}
\usepackage{ulem} 
\usepackage{hyperref}
\usepackage[T1]{fontenc}
\usepackage{CJKutf8} 

\DeclareRobustCommand{\VAN}[3]{#2}
\let\VANthebibliography\thebibliography
\def\thebibliography{\DeclareRobustCommand{\VAN}[3]{##3}\VANthebibliography}

\usepackage{graphicx}	
\usepackage{amsmath}	
\usepackage{amssymb}	

\usepackage[utf8]{inputenc}
\usepackage{siunitx}
\usepackage[dvipsnames]{xcolor} 
\usepackage{float}
\usepackage{soul}

\newcommand\id{{\mathrm d}}

\title[Constraining the EOS with radio pulsars]{Constraining the dense matter equation-of-state with radio pulsars}

\author[H. Hu et al.]{Huanchen Hu (胡奂晨)$^{1}$\thanks{E-mail: huhu@mpifr-bonn.mpg.de},
    Michael Kramer$^{1,2}$,
    Norbert Wex$^{1}$,
    David J. Champion$^{1}$,
    \newauthor
    and Marcel S. Kehl$^{1}$
\\
$^{1}$Max-Planck-Institut f\"ur Radioastronomie, Auf dem H\"ugel 69, D-53121 Bonn, Germany\\
$^{2}$Jodrell Bank Centre for Astrophysics, The University of Manchester, Oxford Road, Manchester M13 9PL, United Kingdom
}
\date{Accepted 2020 July 15. Received 2020 July 15; in original form 2020 June 23}

\pubyear{2020}

\begin{document}
\label{firstpage}
\pagerange{\pageref{firstpage}--\pageref{lastpage}}

\begin{CJK*}{UTF8}{gkai}
\maketitle
\end{CJK*}

\begin{abstract}
Radio pulsars provide some of the most important constraints for our understanding of matter at supranuclear densities. So far, these constraints are mostly given by precision mass measurements of neutron stars (NS). By combining single measurements of the two most massive pulsars, J0348$+$0432 and J0740$+$6620, the resulting lower limit of 1.98\,$M_\odot$ (99\% confidence) of the maximum NS mass, excludes a large number of equations of state (EOSs). 
Further EOS constraints, complementary to other methods, are likely to come from the measurement of the moment of inertia (MOI) of binary pulsars in relativistic orbits.
The Double Pulsar, PSR~J0737$-$3039A/B, is the most promising system for the first measurement of the MOI via pulsar timing. 
Reviewing this method, based in particular on the first MeerKAT observations of the Double Pulsar, we provide well-founded projections into the future by simulating timing observations with MeerKAT and the SKA. 
For the first time, we account for the spin-down mass loss in the analysis. Our results suggest that an MOI measurement with 11\% accuracy (68\% confidence) is possible by 2030. 
If by 2030 the EOS is sufficiently well known, however, we find that the Double Pulsar will allow for a 7\% test of Lense-Thirring precession, or alternatively provide a $\sim3\sigma$-measurement of the next-to-leading order gravitational wave damping in GR. 
Finally, we demonstrate that potential new discoveries of double NS systems with orbital periods shorter than that of the Double Pulsar promise significant improvements in these measurements and the constraints on NS matter.

\end{abstract}


\section{Introduction}
\label{sec:intro}

Neutron stars (NSs) are among the most compact and exotic objects in nature, comprised of extraordinarily dense matter that is not accessible in laboratory experiments. Determining the properties and structure of the cold dense matter inside NSs is therefore a tremendous challenge in nuclear physics. Thus far, a variety of equations of state (EOSs) have been proposed to describe the pressure – density relation inside NSs \citep[see e.g.,][]{Lattimer_2001,Lattimer_2016}. Constraining the EOS is crucial for understanding aspects of fundamental physics, such as the internal structure of NSs, the dynamics of binary mergers, and r-process nucleosynthesis \citep[for a recent review see][]{OF2016}.

Various observational methods have emerged to measure the macroscopic properties of NSs, which promise to increase our knowledge of the EOS. The gravitational wave (GW) observation of a binary NS merger with LIGO/Virgo offers the possibility of measuring the tidal deformability \citep{LIGO2017, LIGO2018}. X-ray observations of emissions from the hot regions on NS surface with NICER \citep{Watts2016} allows a joint mass-radius estimation \citep{Riley2019, Miller2019}.

The largest number of known NSs, however, can be observed as radio pulsars. Currently about 3000 pulsars are known, and the ability of radio astronomers to measure pulsar properties precisely via a technique known as ``pulsar timing'', suggests that important information about the EOS of NSs can also be derived from such measurements. This is indeed the case. The most direct and best known route is to measure the masses of NSs precisely. This is possible in binary pulsars using relativistic orbital effects, potentially combined with other information. The mass range, especially the maximum mass observed, must obviously be consistent with the range of masses supported by a proposed EOS. In addition, there are other orbital effects that also offer the possibility of measuring the moment of inertia (MOI) in binary pulsars via relativistic spin-orbit coupling, as was first suggested by \cite{DS88}. The MOI of a NS depends crucially on the EOS and hence allows us to constrain or even identify it \citep{Morrison2004, LS05, Greif_2020}.  Accessing the MOI of isolated NSs, in contrast, may be possible if one can reliably derive or measure the total loss in rotational energy, $\dot{E}$, which relates the MOI with the observed period and period derivative.

In this work, we provide insight into the
various methods using binary pulsars and their current status in Section~\ref{sec:methods}, before we focus specifically on the possibility of using the Double Pulsar \citep{Burgay2003,Lyne2004} for MOI measurements. 
We will provide an in-depth study of the relevant factors in Section~\ref{sec:LT_binary}, where we explain how Lense-Thirring (LT) precession affects the periastron advance. Section~\ref{sec:pbdot} describes the intrinsic and extrinsic contributions to the orbital period decay. We describe how we simulate future timing observations in Section~\ref{sec:sim} and evaluate how the Double Pulsar can measure the MOI and constrain the EOS in Section~\ref{sec:MOI}. Prospects of testing LT precession and constraining theories of gravity is discussed in Section~\ref{sec:LT} by assuming the EOS is known. We investigate potential constraints on next-to-leading order GW damping in Section~\ref{sec:3p5}, and potential constraints from future discoveries of more relativistic binary pulsars in Section~\ref{sec:new}. Finally, we conclude in Section~\ref{sec:dis}.

\section{Methods to constrain the EOS via pulsar timing}
\label{sec:methods}
 
\subsection{Mass measurements}

A given EOS $i$ can only sustain a NS up to a certain maximum mass, $M^{\rm max}_i$. Finding a massive NS of mass $M_j$, consequently excludes all EOS with $M^{\rm max}_i < M_j$. This was, for instance possible, by using a Shapiro delay measurement in PSR~J1614$-$2230, where \cite{dpr+10} determined a mass $M=1.97\pm 0.04 \,M_\odot$. We note that a recent update on continued timing observations \citep{nanograv11}, implies a significantly lower mass of $1.908 \pm 0.016\,M_\odot$ for this pulsar. As pointed out by \cite{Cromartie_2019}, a Shapiro delay measurement and the determined uncertainty can be affected by the exact orbital sampling (see also Hu et al. in prep.).

In 2013, \cite{Antoniadis_2013} could determine the mass of PSR~J0348$+$0432 without using a Shapiro delay measurement. They combined radio timing measurements of the orbit of the pulsar with precise spectroscopy data of the white dwarf companion in the optical regime to derive a mass of $2.01 \pm 0.04\,M_\odot$, confirming the existence of 2-$M_\odot$ NSs via a complementary method. 

\begin{figure}
    \includegraphics[width=\columnwidth]{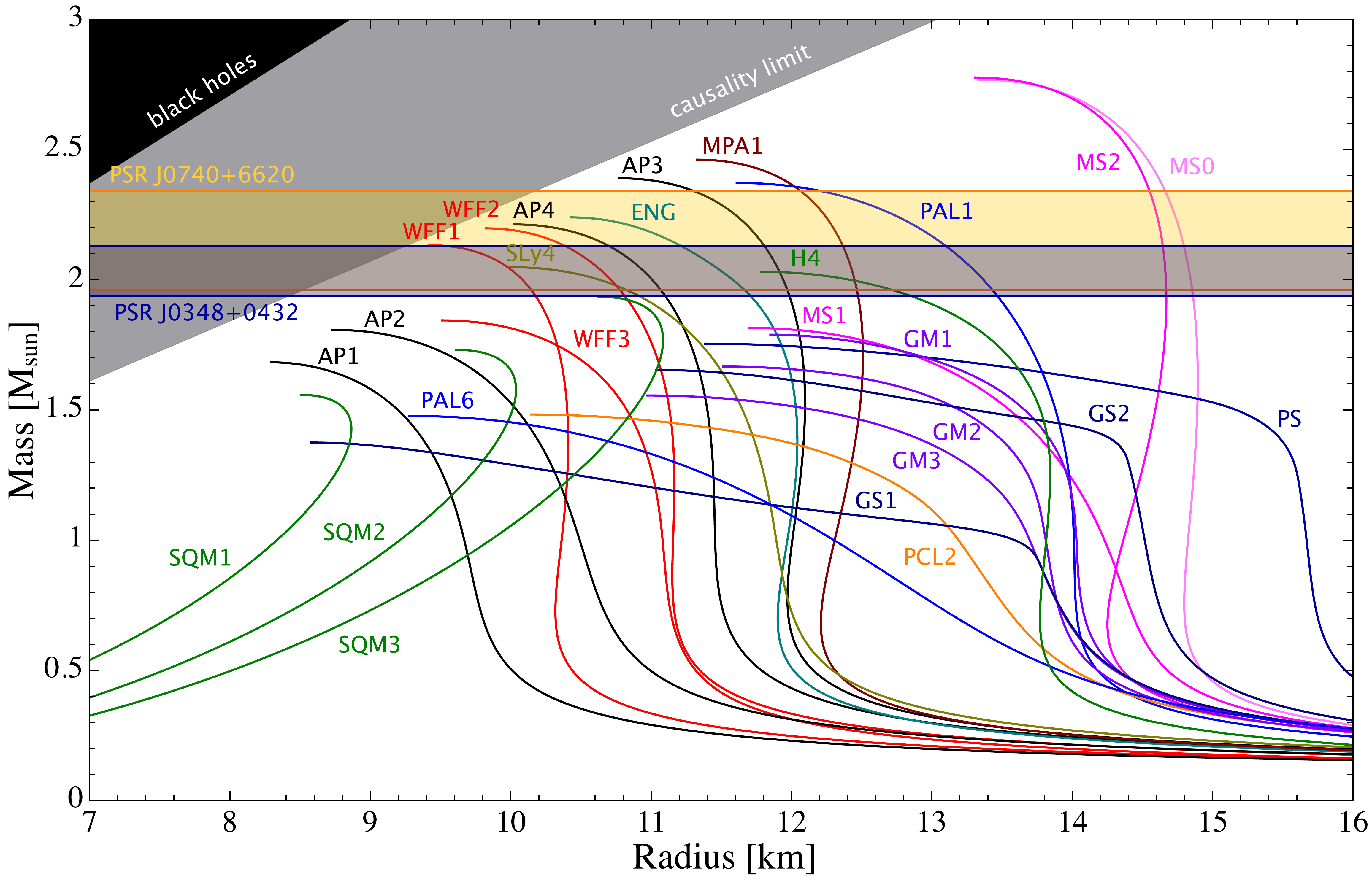}
    \caption{The mass of a NS as function of its radius for different EOS \citep{Lattimer_2001}. The horizontal bands indicate the 2-$\sigma$ range for the masses of the two most massive radio pulsars known to date, PSR J0348$+$0432 \citep{Antoniadis_2013} in blue and PSR J0740$+$6620 \citep{Cromartie_2019} in yellow.}
    \label{fig:R-M}
\end{figure}

Recently, \cite{Cromartie_2019} used a Shapiro delay measurement in PSR~J0740$+$6620 to determine a pulsar mass of $2.14_{-0.09}^{+0.10}\,M_\odot$. We can use the masses of these latter two most massive pulsars, J0348$+$0432 (fully accounting for the rather asymmetric probability density distribution found by \cite{Antoniadis_2013}) and J0740$+$6620, to obtain a 99\% confidence lower limit for the maximum mass of a NS, $1.98\,M_\odot < M^{\rm max}$. Such a constraint already rules out a number of soft EOSs as shown in Figure~\ref{fig:R-M}.\footnote{Note that in Fig.~\ref{fig:M-I}, we show a different but overlapping set of EOSs. Here, we also show EOSs that have been excluded by the maximum mass measurement, while at the same time making the plot not too crowded.}

We can compare this lower limit derived from pulsar timing with an upper limit placed by the NS-NS merger GW170817 observed by LIGO \citep{LIGO2017}. Assuming that the NS-NS merger resulted in the formation of a black hole, one finds an upper limit of about $2.3\,M_\odot$ for the maximum mass of a NS \citep{Rezzolla_2018,Shibata_2019}.

\subsection{Relativistic spin-orbit coupling}
\label{sec:methodSO}

Unlike in Newtonian gravity, the gravitational field of a body in general relativity (GR) has contributions from the mass currents related to the body's proper rotation. \cite{Lense_1918} --- with substantial help from Albert Einstein \citep[see][]{Pfister_2007} --- have shown that the rotation of the Sun has, in principle, an effect on the planetary orbits. This relativistic spin-orbit coupling, also known as LT precession, has since been well tested in the gravitational field of the rotating Earth with the help of satellite laser ranging \citep{Ciufolini_2004,Ciufolini_2019}. Similarly, in relativistic binaries, the spin of a compact rotating body is expected to couple gravitationally with the orbital motion of the system \citep{Barker1975}, leading to a precession of the orbit, while the total angular momentum is conserved.\footnote{The loss of orbital angular momentum due to the emission of GWs is of higher post-Newtonian order and can be neglected here.} This LT precession of the orbit is potentially observable, hence providing a route to access the MOI of the pulsar \citep{DS88}. An MOI measurement, even with an accuracy of $\sim$10\%, would offer important constraints of the EOS \citep{Morrison2004, LS05}. 

The LT precession of the orbit may be detected via the variation in the orbital inclination angle, $i$, and hence in the (observable) projected semi-major axis of the pulsar obit, $x=a_\mathrm{p} \sin i/c$ ($a_\mathrm{p}$ is the semi-major axis, and $c$ the speed of light). However, for this to be detectable, the misalignment angle between pulsar spin and angular momentum vector must be sufficiently large. Also, the orbital inclination angle must not be too close to 90 degrees (``edge-on'' geometry), since the precession leads to a contribution to the rate of change of the projected semi-major axis given by 
\begin{equation}\label{eqn: xdot}
\dot{x}^\text{LT} = x\cot i\left(\frac{\text{d}i}{\text{d}t}\right)_\text{LT},
\end{equation}
where $(\text{d}i/\text{d}t)_\text{LT}$ is given by Eq.(3.27) in \cite{DT92}. For nearly edge-on systems, i.e. $i\approx 90$ deg, this contribution becomes small and most likely undetectable since 
$\cot i \ll 1$. However, in relativistic binary systems with smaller inclination angles, such as PSR J1757$-$1854, this measurement appears to be possible by \cite{Cameron2018}. To achieve this, two challenges will have to be addressed successfully. Firstly, the precession is expected to cause a variation in the pulse profile with time due to a change in the viewing geometry \citep[e.g.][]{Kramer_1998}. Special care in the timing procedure is then needed to obtain sufficient precision and to properly account for possible systematic errors \citep[e.g.][]{stairs02,bhat08,leeuwen1906}. Moreover, since we require access to the direction of the pulsar spin vector \citep{DS88,DT92}, the geometry of the binary system and the pulsar needs also to be measured. This is, however, possible via polarisation measurements, as shown previously \citep[e.g.][]{Kramer_1998,stairs04,gregory1906,vivek_Science}.

Alternatively, rather than using a contribution to $\dot{x}$, one can exploit LT precession also via its contribution to the advance of periastron. If it is possible to isolate the contribution of $\dot{\omega}^\text{LT}$ from the total periastron advance, then the MOI can be determined \citep{DS88}.
This method was suggested for the Double Pulsar, PSR~J0737$-$3039A/B \citep{Lyne2004, Kramer2006}. \cite{KW2009} concluded that a MOI measurement of $\sim$10\% accuracy is possible by $\sim$2030, with the timing accuracy achievable at the time. Later, \cite{Kehl} simulated timing data from emerging telescopes, i.e., the Square Kilometer Array \citep[SKA; e.g.][]{Kramer2015} and its precursor MeerKAT \citep{camilo2018,bailes2018}, which greatly improve the timing precision, and predict a MOI measurement with an accuracy well below 10\% by 2030. However, the timeline of the nominal operation of the SKA assumed by \cite{Kehl} was optimistic compared to the current estimates. With MeerKAT in operation since about 2018 (albeit initially with limited capability), operations of the first phase of the SKA (SKA 1, initially expected to have about 10\% of the full SKA's sensitivity) are not expected before 2027. However, first useful data from commissioning observations may be already available in 2025. \footnote{See \href{www.skatelescope.org}{skatelescope.org} for updates.} 
In addition, compared to \cite{Kehl}, we now already have about two years of Double Pulsar timing observations with MeerKAT, and therefore have more realistic numbers for the timing precision and cadence of observations, not only for the current MeerKAT configuration but also for future extensions.
Moreover, \cite{Kehl} did not incorporate the contribution of spin-down mass loss of pulsar A to the orbital period derivative into the simulations. As we will show below, considering this effect is important, and its impact on our ability to measure the MOI needs to be studied in a fully consistent analysis. Hence, more complete simulations of the MOI measurement in the Double Pulsar should give us a more realistic estimate of the system's (near) future capability to constrain the EOS of ultra-dense matter inside a NS.

Consequently, in what follows, we present new and important details of how to measure the MOI of radio pulsars using the method of isolating the LT contribution to the advance of periastron. Using the Double Pulsar as the most promising system for this kind of experiment, we simulate timing data of PSR~J0737$-$3039A that can be expected from MeerKAT and future extensions, to assess our ability to measure its MOI in the next 10 years. 

\section{Lense-Thirring effect in the Double Pulsar}
\label{sec:LT_binary}
The Double Pulsar is the only system to-date where both NSs have been observed as pulsars \citep{Burgay2003, Lyne2004}, with an orbital period of only 2.4~h. \cite{Breton2008} used the system to provide a 13\%-test of spin-orbit interaction of strongly self-gravitating bodies using the relativistic spin precession in pulsar B. The compact, relativistic nature of the system also allows the measurement of several post-Keplerian (PK) parameters to an unparalleled level of accuracy.
This not only enables some of the most stringent tests of GR related to strong-field gravity \citep{Kramer2006,KW2009,Will_2018}, but it is also crucial for the efforts to measure the MOI and to constrain the EOS of a NS.
\subsection{Spin-orbit coupling contribution to the periastron advance}
\label{subsec:SO}

To simplify the problem, we neglect the LT contribution of pulsar B, since it spins about 122 times slower than pulsar A. Such a simplification is well justified, as will become clear below. In addition, the long term observations of the pulse profile of PSR~J0737$-$3039A shows that the misalignment angle between the spin axis of pulsar A and the orbital angular momentum has an upper limit of $3.2\si{\degree}$ \citep{Ferdman2008, Ferdman2013}. Therefore, for all practical purposes, we can assume that the spin of pulsar A is parallel to the orbital angular momentum, which is consistent with evolutionary considerations for the Double Pulsar system and a low-kick supernova formation \citep[e.g.][]{0737evolv,tauris17}. \cite{pol2018} confirmed that pulsar A is indeed rotating prograde in its orbit, using the modulation of pulsar B's radio emission by the interaction with the wind of pulsar A.
Consequently, the spin of pulsar A only induces a change to the advance of periastron, and does not lead to a change in the orbital inclination, more specifically, the projected semi-major axis. Following \cite{DS88}, the total intrinsic contribution to the periastron advance in the Double Pulsar system can be written, with sufficient precision, as
\begin{align}
\dot{\omega}^\mathrm{intr} & = \dot{\omega}^\text{1PN}  + \dot{\omega}^\text{2PN} + \dot{\omega}^\text{LT,A} \,\notag\\
&= \frac{3\,\beta_\mathrm{O}^2\, n_\mathrm{b}}{1-e_\mathrm{T}^2}\left[1 + f_\mathrm{O}\, \beta_\mathrm{O}^2 - g_\mathrm{S_A}^\parallel \beta_\mathrm{O}\,\beta_\mathrm{S_A}\right] \,,
\label{eq:omdot}
\end{align}
where $n_\mathrm{b}$ is the orbital frequency, and $e_\mathrm{T}$ is the proper-time eccentricity used as the observed eccentricity in the standard timing model \citep{TEMPO:2015} and defined in \cite{DD86}. The factor in front of the right-hand side of Eq.~\eqref{eq:omdot} is the first post-Newtonian (1PN) contribution; the higher order corrections due to 2PN effects and LT precession caused by pulsar A are indicated by the second and third term in the square brackets respectively. The following notations are used to simplify Eq.~\eqref{eq:omdot}, 
\begin{align}
\beta_\mathrm{O} &= \frac{(G M n_\mathrm{b})^{1/3}}{c} \,,\\
\beta_{\mathrm{S}_\mathrm{A}} &=\frac{c I_\mathrm{A} \Omega_\mathrm{A}}{Gm_\mathrm{A}^2} \,,\\
f_\mathrm{O} &= \frac{1}{1-e_\mathrm{T}^2} \left(\frac{3}{2}x_\mathrm{A}^2 + \frac{3}{2}x_\mathrm{A} + \frac{27}{4} \right) +\left(\frac{5}{6}x_\mathrm{A}^2 - \frac{23}{6}x_\mathrm{A} - \frac{1}{4} \right),\\
g_\mathrm{S_A}^\parallel &= \frac{1}{(1 - e_\mathrm{T}^2)^{1/2}} \left(\frac{1}{3} x_\mathrm{A}^2 + x_\mathrm{A} \right)\,.
\end{align}
The subscript A stands for pulsar A. $G$ is the Newtonian gravitational constant, and $M = m_\mathrm{A} + m_\mathrm{B}$ is the total mass defined as the sum of the (inertial) masses of pulsar A and B, and $x_\mathrm{A} = m_\mathrm{A}/M$. $I_\mathrm{A}$ denotes the MOI and $\Omega_\mathrm{A}$ the angular spin frequency. \footnote{Since pulsar A is slowly rotating ($\sim$ 2.5\% of break-up velocity), for the purpose of this paper we do not have to distinguish between rotating and non-rotating quantities when it comes to (gravitational) mass, moment of inertia, etc. \protect\citep[see e.g.][]{Berti_2005}}

Table~\ref{tab:omdot} lists the values of each term contributing to $\dot{\omega}^\mathrm{intr}$, using the Keplerian parameters and masses ($m_\mathrm{A}= 1.3381\, M_\odot$, $m_\mathrm{B}=1.2489\, M_\odot$) measured in \cite{Kramer2006}. We note that the contribution due to the LT precession $\dot{\omega}^\text{LT,A}$ depends on the MOI, whereby is written as a function of $I_\mathrm{A}^{(45)}$ defined as $I_\mathrm{A}^{(45)}=I_\mathrm{A} / (10^{45}\,\mathrm{g\,cm^2})$. Typical values of $I_\mathrm{A}^{(45)}$ are around unity for realistic EOSs. It is evident that the contribution from the LT effect is comparable to that of 2PN, but with opposite signs. 

The analysis of timing data from relativistic binary pulsars is based on a particularly simple and elegant solution of the post-Newtonian equations of motion, the so called Damour-Deruelle (DD) model \citep{DD85,DD86,TEMPO:2015}. In the quasi-Keplerian parametrization of the DD model one can see that the advance of periastron is proportional to the true anomaly. This behaviour is modified by two periodic terms as part of the generalised quasi-Keplerian parametrization, which is a natural extension of the DD model when including 2PN and spin-orbit terms \citep{DS88,SW93,Wex1995}. However, these periodic terms will remain well below measurability for the foreseeable future, for any of the known binary pulsars. For that reason, we will ignore them in our analysis.

Besides the coupling to the orbital angular momentum (spin-orbit coupling), the spin of pulsar A also couples to the spin of pulsar B (spin-spin coupling) \citep{Barker1975}. However, the spin of pulsar B is about a factor of $3 \times 10^6$ smaller than the orbital angular momentum. Hence, spin-spin coupling is totally irrelevant here.

Finally there are, at least in principle, also contributions from the rotationally induced mass quadrupole moments of pulsars A and B to the orbital dynamics \citep{Barker1975}. These spin-squared contributions give rise to an additional change in the advance of periastron \citep{Smarr1976,Wex1998}. The contribution from the quadrupole moment of pulsar A is estimated to be $\sim 3 \times 10^{-8}$ deg\,yr$^{-1}$, where we have used the relations in \cite{Bauboeck2013} to calculate the mass quadrupole. This is four orders of magnitude smaller than the second order effects. The contribution from pulsar B is even smaller (about $10^4$ times) due to its slower rotation. Hence we can totally ignore such contributions in this study.

{\renewcommand{\arraystretch}{1.5}
\begin{table}
\caption{Contributions to the rate of periastron advance in the Double Pulsar calculated using Eq.~\eqref{eq:omdot}, with the Keplerian parameters and masses ($m_\mathrm{A}= 1.3381\, M_\odot$, $m_\mathrm{B}=1.2489\, M_\odot$) measured in \citet{Kramer2006}. $I_\mathrm{A}^{(45)}=I_\mathrm{A} / (10^{45}\mathrm{g\,cm^2})$. The current measurement precision for $\dot\omega$ is already $\sim$$10^{-5}\,$deg\,yr$^{-1}$ (Kramer et al. in prep.), which is about 40 times smaller than $\dot{\omega}^\text{LT,A}$.
}
\begin{center}
\begin{tabular}{lrl}
\hline \hline
Contribution   &  [deg\,yr$^{-1}$] \\ \hline
$\dot{\omega}^\text{1PN}$ & 16.898703 \\ 
$\dot{\omega}^\text{2PN}$ & \;\;0.000439  \\ 
$\dot{\omega}^\text{LT,A}$  & $-0.000377$ & \hspace*{-1.4em} $\times\, I_\mathrm{A}^{(45)}$ \\ \hline
\end{tabular}
\end{center}

\label{tab:omdot}
\end{table}
}

\subsection{The proper motion contribution to the observed periastron precession}

Apart from the intrinsic contributions to the periastron advance, the proper motion of a binary system also can change the apparent geometrical orientation of the orbit, and hence the observed periastron advance \citep{Kopeikin1996}. As a consequence, the observed value of periastron advance is shifted from its intrinsic value by
\begin{align}
\dot{\omega}^\mathrm{obs} = \dot{\omega}^\mathrm{intr} + \dot{\omega}^\mathrm{K} \,.
\label{eq:omdot_intr}
\end{align}
Here, $\dot{\omega}^\mathrm{K}$ is the {\it Kopeikin term} that satisfies
\begin{align}
    \dot{\omega}^\mathrm{K} = 2.78 \times 10^{-7} \csc i\, (\mu_\alpha \cos\Omega + \mu_\delta \sin \Omega) \;\mathrm{deg\,yr^{-1}} \,,
\end{align}
where $i$ is the orbital inclination as defined in \citet{DT92}, $\mu_\alpha$ and $\mu_\delta$ are the proper motion in right ascension and declination, and $\Omega$ is the longitude of the ascending node (measured from East, in the sense of rotation towards North). Using the parameters measured by \cite{Kramer2006} and the estimated $\Omega=25(2) \si{\degree}$ by \cite{Rickett2014}\footnote{Note, \cite{Rickett2014} use a different definition for the longitude of the ascending node $\Omega$, which we have accounted for.}, we obtain $\dot{\omega}^\mathrm{K}=-4.0(3) \times 10^{-7}\,\mathrm{deg\,yr^{-1}}$. Given the current measurement precision $\Delta \dot{\omega} \sim 10^{-5}\, \mathrm{deg\,yr^{-1}}$ (Kramer et al. in prep.), the {\it Kopeikin term} is a small correction to the intrinsic periastron advance that we use in this study. However, since it is three orders of magnitude smaller than $\dot{\omega}^\text{LT,A}$ (see Table~\ref{tab:omdot}), it does not have a significant influence on the LT measurement.

\subsection{Challenges on extracting the Lense-Thirring contribution and measuring the MOI}

Although the current measurement precision $\Delta\dot{\omega}$ is already $\sim$ 40 times smaller than $\dot{\omega}^\text{LT,A}$, it is not that straightforward to extract the LT contribution from $\dot{\omega}^\mathrm{obs}$ with Eqs.~\eqref{eq:omdot} and \eqref{eq:omdot_intr}, as the two masses ($m_\mathrm{A},\,m_\mathrm{B}$) are needed to calculate $\dot{\omega}^\text{1PN}$ and $\dot{\omega}^\text{2PN}$. The masses need to be obtained from any other two PK parameters, where the best two here are the Shapiro delay shape parameter $s$ and the orbital period derivative $\dot{P}_\mathrm{b}$ (see Figure~\ref{fig:frac_error}). For the Double Pulsar, we already have sufficient precision for $s$, so the limitation is mainly from $\dot{P}_\mathrm{b}$ (Kramer et al. in prep.). The measurement precision of $\dot{P}_\mathrm{b}$ will improve over time, especially with the addition of MeerKAT and the SKA. However, the observed value of $\dot{P}_\mathrm{b}$ is influenced by extrinsic acceleration effects, which depend on the distance of the pulsar and the Galactic gravitational potential. Moreover, the spin-down mass loss of the pulsars also have an impact on $\dot{P}_\mathrm{b}$, which itself depends on the MOI, meaning the masses can not be determined independently from $I_\mathrm{A}$. These are the challenges for measuring the MOI. An alternative option to $\dot{P}_\mathrm{b}$ could be the time dilation amplitude $\gamma$, whose fractional error is about one order of magnitude larger than $\dot{P}_\mathrm{b}$ (see Figure~\ref{fig:frac_error}). However, based on the assumptions of observing plan in Section~\ref{sec:sim}, it would take at least two decades from now to obtain a 1$\sigma$-measurement of $\dot{\omega}^\text{LT,A}$ using only $\gamma$, $s$ and $\dot{\omega}$, a precision that is already reached now with $\dot{P}_\mathrm{b}$ (Kramer et al. in prep.). Hence, a comprehensive understanding of the individual contributions to $\dot{P}_\mathrm{b}$ is needed, which will be discussed in detail in the following section.

\section{The intrinsic and extrinsic contributions to the orbital period decay}
\label{sec:pbdot}

The observed value of the orbital period decay comprises several effects \citep{DT91}. For the purpose of this study, we only consider the dominant terms
\begin{equation}
\left(\frac{\dot{P}_\mathrm{b}}{P_\mathrm{b}}\right)^\text{obs} 
= \left(\frac{\dot{P}_\mathrm{b}}{P_\mathrm{b}}\right)^{\text{GR}}
+ \left(\frac{\dot{P}_\mathrm{b}}{P_\mathrm{b}}\right)^{\dot{m}_\mathrm{A}}
+ \left(\frac{\dot{P}_\mathrm{b}}{P_\mathrm{b}}\right)^{\text{Gal}} 
+ \left(\frac{\dot{P}_\mathrm{b}}{P_\mathrm{b}}\right)^{\text{Shk}} 
 \,,
\label{eq:pbdot}
\end{equation}
where gravitational wave damping (GR) and mass loss of pulsar A ($\dot{m}_\mathrm{A}$) are intrinsic contributions, and Galactic acceleration (Gal) and Shklovskii effect (Shk) are extrinsic contributions. Thereby, the intrinsic orbital period decay can be extracted from the observed value using
\begin{align}
    \dot{P}_\mathrm{b}^\text{ intr}
    & = \dot{P}_\mathrm{b}^\text{ obs} -  \dot{P}_\mathrm{b}^{\text{ Gal}} - \dot{P}_\mathrm{b}^{\text{ Shk}} .
    \label{eq:intr}
\end{align}
Consequently, the uncertainty in the intrinsic orbital period decay also depends on the error in the pulsar distance and the uncertainty in the Galactic gravitational potential at the location of the pulsar and the Earth.


\subsection{Gravitational wave damping}

The binary system loses energy in the form of GW emission, which shrinks the orbit of the system, and in turn gradually reduces the orbital period. The post-Newtonian approximation is employed to describe the orbital dynamics of the binary system \citep[see e.g.][]{Damour_1987,Blanchet_LRR}, i.e.\ the equations of motion are expanded with respect to $v/c$, where $v$ denotes a typical orbital velocity. The  change of the orbital period due to GW damping enters at order $(v/c)^5$, i.e.\ the 2.5PN approximation. The corresponding change in the orbital period is given by \citep{Peters1963, Esposito1975, Wagoner1975}
\begin{align}
\dot{P}_\mathrm{b}^{\text{ 2.5PN}}= -\frac{192\pi}{5} \frac{\eta\,\beta_\mathrm{O}^{\,5}}{(1-e_\mathrm{T}^2)^{7/2}}\left( 1+\frac{73}{24}e_\mathrm{T}^2+ \frac{37}{96}e_\mathrm{T}^4\right) \,,
\label{eq:pbdot_2p5PN}
\end{align}
where $\eta = m_\mathrm{A} m_\mathrm{B} /M^2$ is the symmetric mass ratio. Later, \cite{BS89} extended the expression to the next-to-leading order (3.5PN), 
\begin{align}
\dot{P}_\mathrm{b}^{\text{ GR}}= &-\frac{192\pi}{5} \frac{\eta\,\beta_\mathrm{O}^{\,5}}{(1-e_\mathrm{T}^2)^{7/2}} \Bigg\{ 1+\frac{73}{24}e_\mathrm{T}^2+ \frac{37}{96}e_\mathrm{T}^4 \notag\\
&+ \frac{\beta_\mathrm{O}^{\,2}}{336\, (1-e_\mathrm{T}^2)}
\Bigg[1273 + \frac{16495}{2}e_\mathrm{T}^2 + \frac{42231}{8}e_\mathrm{T}^4 + \frac{3947}{16}e_\mathrm{T}^6 \notag\\
&- \left(924 + 3381 e_\mathrm{T}^2+ \frac{1659}{4}e_\mathrm{T}^4 -\frac{259}{4} e_\mathrm{T}^6 \right) \eta \notag\\
&+ \left(3297 e_\mathrm{T}^2 +4221 e_\mathrm{T}^4 + \frac{2331}{8} e_\mathrm{T}^6 \right) \frac{\delta m}{M} \Bigg] \Bigg\}, \label{eq:pbdot_3p5PN}
\end{align}
where $\delta m$ denotes the mass difference of the timed pulsar and its companion, in our case, $\delta m = m_\mathrm{A} - m_\mathrm{B}$. Eq.~\eqref{eq:pbdot_3p5PN} can be written in a simplified form as
\begin{align}
    \dot{P}_\mathrm{b}^{\text{ GR}}= \dot{P}_\mathrm{b}^{\text{ 2.5PN}} \left(1+ X_\mathrm{3.5PN} \right)\,,
\label{eq:pbdot_GW}
\end{align}
where the relative correction of the 3.5PN order, $X_\mathrm{3.5PN}$, is $1.40\times 10^{-5}$ for the Double Pulsar. To date, only the leading order contribution to the orbital period decay is considered in the analysis and interpretation of any of the known binary pulsars. The higher order correction, however, will need to be included in the future, when we reach the necessary timing precision with emerging powerful radio telescopes such as the SKA. 
We will evaluate future measurability of the 3.5PN contribution to $\dot{P}_{\rm b}$ in Section \ref{sec:3p5}.

Besides the damping of the binary period, the emission of GWs in principle has an additional effect on the observed $\dot{P}_\mathrm{b}$. \cite{Junker_1992} have shown that a double NS system with asymmetric masses in an eccentric orbit becomes accelerated due to the GW recoil. Since any acceleration along the line of sight leads to an apparent change in the orbital period \citep{DT91}, the GW recoil at 3.5PN order will also affect the observed orbital period at some level. As \cite{Junker_1992} have pointed out, the recoil acceleration changes its direction with the advance of periastron, in our case on a timescale of about 21 years. However, using Eq.~(103) in \cite{Junker_1992} we find a maximum shift in the observed $\dot{P}_\mathrm{b}$ due to GW recoil of $4.6 \times 10^{-24}$, which is seven orders of magnitude below the current measurement precision.

\subsection{Galactic acceleration and Shklovskii effect}
\label{subsec:Galactic}

The contribution of Galactic acceleration can be calculated with \citep{DT91, Nice1995, Lazaridis09}
\begin{align}
&\left(\frac{\dot{P}_\mathrm{b}}{P_\mathrm{b}}\right)^{\text{Gal}} = -\frac{K_z |\sin{b}|}{c} - \frac{\Theta_0^2}{c R_0} \notag\\
& \times \Bigg\{\cos{l} + \frac{\beta}{\sin^2l + \beta^2} \left[1+ b_0 \left(1-\sqrt{\sin^2l + \beta^2} \right) \right]^2 \Bigg\} \cos{b} \,,
\label{eq:Gal}
\end{align}
where $\beta=(d/R_0)\cos{b}-\cos{l}$. For the Double Pulsar, the Galactic longitude $l$ is $245.2\si{\degree}$ and the Galactic latitude $b$ is $-4.5\si{\degree}$. As for the distance to the Double Pulsar ($d$), the VLBI observations made by \cite{Deller09} implied a distance of $1.15_{-0.16}^{+0.22}$~kpc, whereas the dispersion measure (DM) favours a distance of about $0.52$~kpc \citep{Cordes2002}. We note that new, preliminary timing and VLBI measurements indicate a distance closer to the DM distance (Kramer et al. in prep.). Hence, for our simulation, we consider an intermediate distance of $0.8\,$kpc with a 10\% error. We will see in Section~\ref{sec:MOI}, using a different distance does not have a big influence on our results.
The vertical contribution of the Galactic acceleration $K_z$ for Galactic height $z \equiv |d \, \sin b| \lesssim 1.5$ kpc can be approximated with the expression \citep{HF04,Lazaridis09}
\begin{align}
K_z[10^{-9} \, \mathrm{cm \, s^{-2}}] \simeq 2.27 \, z_\mathrm{{kpc}} + 3.68 \,\left[1-\exp(-4.31 \, z_\mathrm{kpc})\right],
\end{align}
where $z_\mathrm{kpc} \equiv z [{\rm kpc}]$. For $K_z$, we consider a typical error of about $10\%$ \citep{HF04,Zhang_2013}. The Galactic parameters $R_0$ is the distance from the Sun to the Galactic center, and $\Theta_0$ is the Galactic circular velocity at the location of the Sun. In our calculation, we adopt the recent result in \cite{Gravity2019}, where $R_0 = 8.178 \pm 0.026 \,\text{kpc}$\footnote{We note that the latest measurement of $R_0$ shows a $2\sigma$ difference \citep{Gravity2020}, which will not affect our results.} and $\Theta_0 = 236.9 \pm 4.2 \,\mathrm{km\,s^{-1}}$. The slope parameter at the radius of the Sun is defined as \citep{DT91}:
\begin{align}
    b_0 \equiv \left(\frac{R}{v} \frac{\id{v}}{\id{R}}\right)_{R=R_0} \,.
\end{align}
We note this term is often ignored in other studies, as the rotation curve is nearly flat in the vicinity of the Sun. Its uncertainty, however, could be relevant for measuring the MOI, and as such is included in our study. The slope of the Galactic rotation curve at the location of the Sun estimated by \cite{Reid14} is $-0.2 \pm 0.4 \,\mathrm{km\,s^{-1}\, kpc^{-1}}$, corresponding to $b_0=0.007\pm0.014$. Lately, \citet{Eilers2019} found a slope significantly different from zero, i.e. $-1.7\pm 0.1\,\mathrm{km\,s^{-1}\, kpc^{-1}}$ ($b_0=0.0603 \pm 0.0035$), with a systematic uncertainty of $0.46\,\mathrm{km\,s^{-1}\,kpc^{-1}}$. Both results will be employed later in our simulation, but for \citet{Eilers2019} we only consider the statistical error and assume that the systematic error can be well understood in the future.

Besides the Galactic acceleration, there are additional accelerations due to masses in the vicinity of the Sun or the pulsar, primarily giant molecular clouds (GMCs), but also stars, black holes, and other external masses \citep{DT91, Kehl_thesis}. 
These have most likely, if at all, only a small influence on the result, which should not limit our ability of measuring the MOI with a precision lower than 10\% \citep{Kehl_thesis}. 
The influence of these masses mostly depends on the distance to the Double Pulsar, with which we can, for instance, trace and restrict the presence and influence of GMCs using Galactic carbon monoxide (CO) surveys \citep{Neininger1998, Glover2011}.
We expect that a more firmly established distance measurement in the future will allow a refined analysis to confirm our conclusions.

Finally, the transverse motion of a pulsar leads to an apparent change in the orbital period. This is known as the Shklovskii effect \citep{Shk70}, and is given as
\begin{align}
\left(\frac{\dot{P}_\mathrm{b}}{P_\mathrm{b}}\right)^{\text{Shk}} = \frac{\mu^{2}d}{c},\quad \text{with}\quad \mu^2 = \mu_\alpha^2 + \mu_\delta^2\,.
\end{align}

\subsection{Mass loss}

A pulsar looses mass due to its energy emission, which changes the orbital period by \citep{Jeans1924,Jeans1925}
\begin{align}
  \left(\frac{\dot{P_\mathrm{b}}}{P_\mathrm{b}} \right)^{\dot{m}} = -2\,\frac{\dot{m}_\mathrm{A} + \dot{m}_\mathrm{B}}{M}\,.
\end{align}
Although the emission process of pulsars is not fully understood, the mass-energy loss can be calculated (with sufficient precision) from the loss in rotational kinetic energy, i.e., $\dot{E}^\mathrm{\,rot}_j \simeq \dot{m}_j c^2$ \citep{DT91}, where $\dot{E}_j^\mathrm{\,rot} =  I_j \Omega_j\dot\Omega_j$, with $\Omega_j$ the angular velocity of the (proper) rotation of body $j$ ($j$ = A or B), given in terms of the spin period by $\Omega_j =2\pi/P_j$. Hence,
\begin{equation}
\left(\frac{\dot{P_\mathrm{b}}}{P_\mathrm{b}} \right)^{\dot{m}_j} = \frac{8\pi^2 \dot{P}_j I_j}{c^2 M P_j^3} \,.
\label{eq:mAdot}
\end{equation}
Clearly, the mass-loss correction to the rate of orbital period decay also depends on the MOI, and therefore on the EOS. Table~\ref{tab:pbdot} lists the predicted value of each contribution to $\dot{P}_\mathrm{b}^\text{ obs}$, where the mass-loss contributions are written as a function of $I_\mathrm{j}^{(45)}$. The contribution due to the mass loss of pulsar A is one order of magnitude smaller than that of the Galactic acceleration and the Shklovskii effect, and of the same order of magnitude as the current measurement precision (Kramer et al. in prep.), hence must be considered. The mass-loss contribution of pulsar B, however, is nearly four orders of magnitude smaller than that of pulsar A and thus can be safely ignored.

{\renewcommand{\arraystretch}{1.5}
\begin{table}
\caption{Contributions to the rate of orbital period decay in the Double Pulsar, calculated with Keplerian parameters and masses measured in \citet{Kramer2006}. The Galactic acceleration is computed using Galactic measurements by \citet{Gravity2019} and slope in \citet{Reid14}, and a distance of $0.80\,$kpc is assumed. $I_\mathrm{B}^{(45)}$ is defined in the same way as $I_\mathrm{A}^{(45)}$. The current measurement precision for $\dot{P}_\mathrm{b}$ is below 0.1\,fs/s (Kramer et al. in prep.).}
\begin{center}
\begin{tabular}{lr l}
\hline
\hline
Contribution & [fs/s] 
\\ \hline
$\dot{P}_\mathrm{b}^\text{ 2.5PN}$ & $-$1248
\\ 
$\dot{P}_\mathrm{b}^\text{ Gal}$   & $-$0.38 \\ 
$\dot{P}_\mathrm{b}^\text{ Shk}$   &  0.21 \\
$\dot{P}_\mathrm{b}^\text{ 3.5PN}$ & $-$0.017 \\ 
$\dot{P}_\mathrm{b}^{\,\dot{m}_\mathrm{A}}$ & 0.023  & \hspace*{-1.4em} $\times\, I_\mathrm{A}^{(45)}$ \\
$\dot{P}_\mathrm{b}^{\,\dot{m}_\mathrm{B}}$ & 6.3\,$\times\, 10^{-6}$\quad & \hspace*{-1.4em} $\times\, I_\mathrm{B}^{(45)}$\\ \hline
\end{tabular}
\end{center}
\
\label{tab:pbdot}
\end{table}
}

\section{Simulations}
\label{sec:sim}
In order to investigate the capability of measuring the MOI and testing GR with radio pulsars, we developed a simulation framework to generate and analyse time-of-arrivals (TOAs) for binary pulsars.
In this section, we will describe how we simulate TOAs from emerging telescopes for PSR~J0737$-$3039A based on realistic assumptions, and how to measure PK parameters and timing parallax. 

To simulate TOAs of PSR~J0737$-$3039A from current and future telescopes, knowledge of the sensitivity of the telescopes, as well as (realistic) assumptions about a future observing plans are needed. We consider the best telescopes for observing this pulsar, i.e., MeerKAT and its future arrays. Unfortunately, this pulsar is not in the field of view of the Five-hundred-meter Aperture Spherical radio Telescope \citep[FAST;][]{Nan2011}, the largest radio telescope today and in the near future.

MeerKAT is a precursor for the mid-frequency array of the SKA, which comprises 64 dishes, each with a diameter of 13.5\,m. This corresponds to an effective diameter ($\diameter_\mathrm{eff}$) of 108\,m. Regular timing observations for the Double Pulsar started in 2019 as a part of the MeerTIME project \citep{bailes2018}. The MeerKAT extension, hereafter MeerKAT+, is a joint collaboration of the South African Radio Astronomy Observatory (SARAO) and the Max-Planck-Society (MPG) to extend MeerKAT by the addition of 20 SKA-type dishes, each 15\,m in diameter, to MeerKAT. MeerKAT+ is expected to operate from 2022, providing an increase in sensitivity by 50\% (Kramer, priv. comm.) The first phase of the SKA mid-frequency array, SKA 1-mid, is planned to build 112 additional dishes with 15\,m diameter, extending MeerKAT+ further, with first data from 2025 and full operation after 2027. We summarise the observing plans and the effective diameters of these telescopes in Table \ref{tab:sim}.

In order to estimate the TOA uncertainty of each observing phase, we need to consider noise contributions for pulsar A. The TOA uncertainty of pulsar A with real MeerKAT observations at L-band is about $1.06 \,\mu$s for a 5 minutes integration over the full bandwidth \citep{Bailes2020}. Since the system performance of MeerKAT+ and SKA 1-mid are expected to be similar to that of MeerKAT, and the radiometer noise $\sigma_\mathrm{rn}$ reduces in reverse proportional to the effective collection area of the telescope $A_\mathrm{eff}$, we can therefore calculate the radiometer noise using the relation
\begin{equation}
    \sigma_\mathrm{rn}^\mathrm{tel} = \frac{A_\mathrm{eff}^\mathrm{MK}}{A_\mathrm{eff}^\mathrm{tel}} \sigma_\mathrm{rn}^\mathrm{MK} \,,
\end{equation} 
where the superscript ``MK'' stands for MeerKAT. We are not considering noise budgets other than the radiometer noise, because: 1) The phase jitter has not been detected in the current MeerKAT observations and must be rather small \citep[][Hu et al. in prep.]{Bailes2020}. It may become important in the future observing phase as the radiometer noise reduces, but the influence of jitter can potentially be reduced using Bayesian methods \citep{Imgrund2015} or binning and combining the data in orbital phase. 2) The contributions from scintillation and other effects are expected to be one or more orders of magnitude smaller than the radiometer noise of SKA 1-mid, hence are neglected. As a result, in our simulation, we adopt the TOA uncertainties solely based on the radiometer noise estimation for each observing phase, which can be found in Table~\ref{tab:sim}. 
{\renewcommand{\arraystretch}{1.5}
\begin{table}
\caption{Telescope observing plans, effective diameters and TOA uncertainties (L-band, 5 minute integration time) used for simulation. For 2003 -- 2019, $\sigma_{\mathrm{TOA}}$ are based on observations from multiple telescopes, where the best data are from the GBT. Its typical uncertainty at L-band is shown in the table, whereas the TOAs from the UHF band are expected to be 1.25 times better (Kramer et al. in prep.). The TOA uncertainty for MeerKAT is scaled to 5 minute integrations based on real observations \citep{Bailes2020}, and for MeerKAT+ and SKA 1-mid are scaled referring to MeerKAT.
}
\begin{center}
\begin{tabular}{cccc}
\hline \hline
Year       & Telescope & $\diameter_{\mathrm{eff}}$ [m] & $\sigma_{\mathrm{TOA}}$ [$\mu$s] \\ \hline
2003 -- 2019 & GBT  & 100 & 2.5 \\ 
2019 -- 2022 & MeerKAT    & 108 & 1.06 \\ 
2022 -- 2025 & MeerKAT+   & 127 & 0.76 \\ 
2025 -- 2030 & SKA 1-mid  & 203 & 0.30 \\ \hline
\end{tabular}
\end{center}
\label{tab:sim}
\end{table}}

Based on the above assumptions, we generate TOAs of PSR~J0737$-$3039A that mimic observations with MeerKAT, MeerKAT+, and SKA 1-mid from 2019 to 2030 covering two full orbits per month ($\sim$5 h), and combine them with the existing TOAs from multiple telescopes (Kramer et al. in prep.) to form a long-range dataset (2003--2030). Technically speaking, we only use the observing cadence and TOA uncertainties from the existing TOAs, since the data analysis by Kramer et. al (in prep.) is still ongoing, and in the next steps all TOAs will be simulated to fit our model, under the assumption of Gaussian white noise.

The first step is to create a parameter file (model) for pulsar A. For this, we take precisely measured masses from \cite{Kramer2006}, $m_\mathrm{A}= 1.3381\, M_\odot$, $m_\mathrm{B}=1.2489\, M_\odot$, and assume EOS AP4 \cite[see][]{Lattimer_2001}. This particular choice of EOS satisfies the current lower limit of $1.98\,M_\odot$ (99\% confidence level, hereafter C.L.) for the maximum mass of a NS (see Section~\ref{sec:intro} for details), and also lies in the MOI ranges obtained for pulsar A by \cite{Gorda_2016, Lim2019, Greif_2020}.
The MOI of pulsar A, under this assumption, is therefore $I_\mathrm{A}^{\mathrm{AP4}} = 1.24 \times 10^{45} \mathrm{g\,cm^2}$. We create a parameter file by taking the well measured Keplerian parameters of the Double Pulsar \citep{Kramer2006} and the PK parameters computed from $m_\mathrm{A}$, $m_\mathrm{B}$, and $I_\mathrm{A}$. For the advance of periastron $\dot{\omega}$, we consider first and second order PN terms and the LT contribution. As for the orbital period decay $\dot{P}_\mathrm{b}$, we consider leading order (2.5PN) GW emission, Galactic acceleration, Shklovskii effect and mass loss in pulsar A. The 3.5PN GW term is only considered in Section~\ref{sec:3p5}.

\begin{figure}
    \includegraphics[width=\columnwidth]{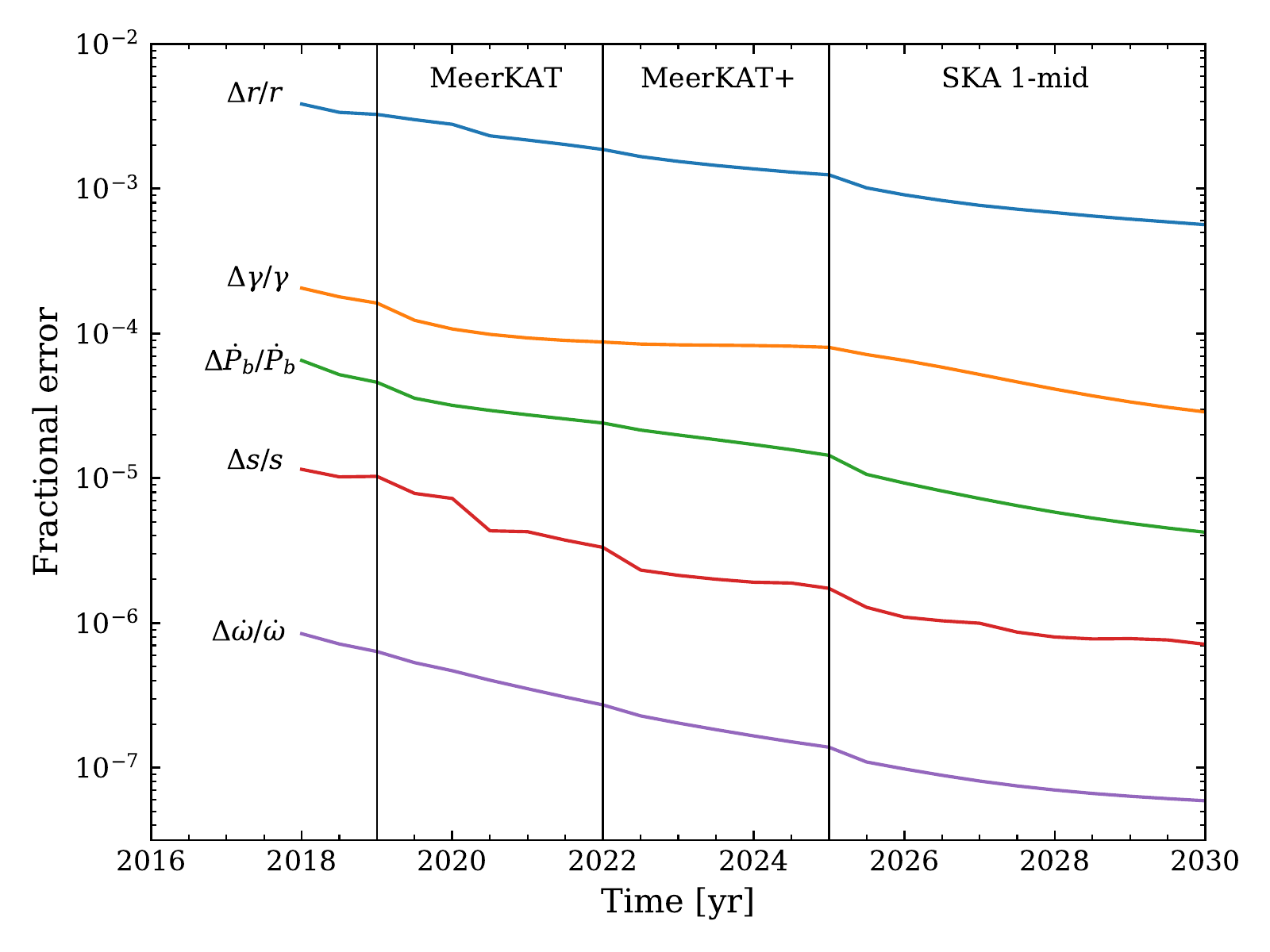}
    \caption{Improvement in the fractional errors of five PK parameters with time, based on the simulation described in Section~\ref{sec:sim}. From top to bottom are: the Shapiro delay range parameter $r$ (blue), the time dilation amplitude $\gamma$ (orange), the orbital period derivative $\dot{P}_\mathrm{b}$ (green), the Shapiro delay shape parameter $s$ (red), and the relativistic advance of periastron $\dot{\omega}$ (purple). The vertical lines mark the observing phase of MeerKAT, MeerKAT+, and SKA 1-mid. }
    \label{fig:frac_error}
\end{figure}

We then adjust the TOAs to perfectly match with our model, and add a Gaussian white noise to each TOA, according to its $\sigma_{\rm TOA}$. The red noise from DM variations is not considered in our simulation, since it can be in principle corrected for with multi-frequency data. In a final step, we use the pulsar timing software TEMPO\footnote{\url{http://tempo.sourceforge.net/}} to fit for the timing parameters and obtain their uncertainties, including the PK parameters, which are of particular importance here. From 2018 to 2030, the dataset is split with a step size of 6 months, so as to demonstrate how the measurements improve with time. The predicted fractional errors of the PK parameters are shown in Figure~\ref{fig:frac_error}.

\begin{figure}
    \includegraphics[width=\columnwidth]{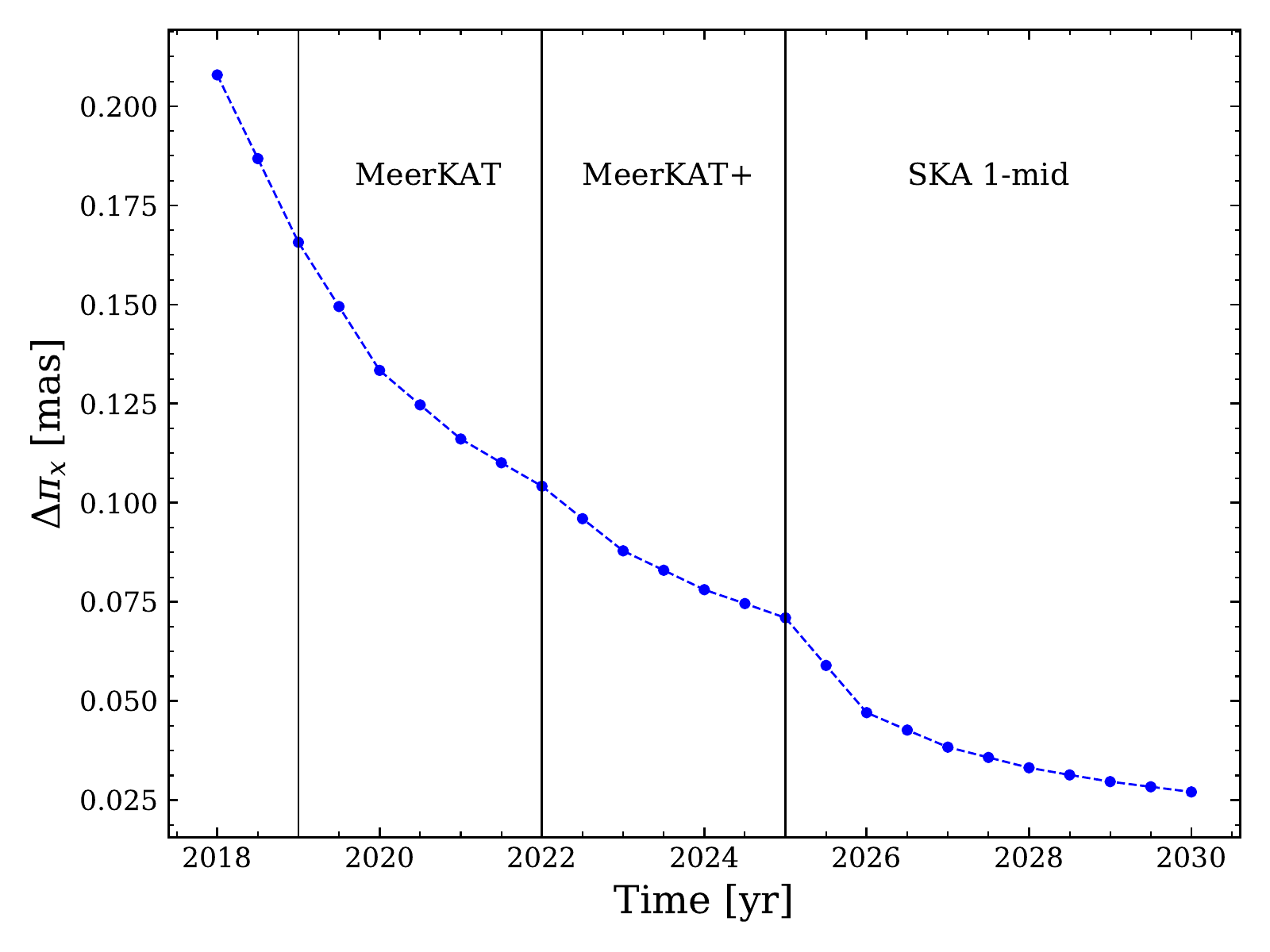}
    \caption{Predicted uncertainty of the timing parallax $\Delta \pi_x$ as a function of time. The corresponding uncertainty in distance is smaller than our assumed value from mid-2021, and is therefore used for future corrections of extrinsic acceleration effects. \label{fig:d_PX} 
    }
\end{figure}

As part of the simulation, we also measure the timing parallax $\pi_x$, which gives an idea of the precision of future distance measurement from timing parallax. The predicted uncertainty of $\pi_x$ is shown in Figure~\ref{fig:d_PX}. For the uncertainty of pulsar distance, which enters the Galactic acceleration and the Shklovskii effect, we adopt the value calculated from timing parallax when its uncertainty is smaller than what we assumed in Section~\ref{subsec:Galactic}, which is from mid-2021. Aside from timing parallax measurement, in the future, the VLBI parallax measurements with the SKA can potentially provide an accurate distance measurement \citep{Smits2011}.

\section{Measuring the MOI and constraining the EOS}
\label{sec:MOI}

Based on our TOA simulation, we predict the future timing measurement of PK parameters (Figure~\ref{fig:frac_error}). The three best measured parameters, $\dot{P}_\mathrm{b}^\text{ obs}$, $\dot{\omega}^\text{obs}$ and $s$, are promising for the determination of $I_\mathrm{A}$. With Eqs.~\eqref{eq:omdot_intr} and \eqref{eq:intr}, we obtain the intrinsic periastron advance $\dot{\omega}^\text{ intr}\,(m_\mathrm{A}, m_\mathrm{B}, I_\mathrm{A})$ and the intrinsic orbital period decay $\dot{P}_\mathrm{b}^\text{ intr}(m_\mathrm{A}, m_\mathrm{B}, I_\mathrm{A})$. Since both now $\dot{\omega}^\text{\,intr}$ and $\dot{P}_\mathrm{b}^\text{ intr}$ depend on the MOI, we can not directly use $\dot{P}_\mathrm{b}^\text{ intr}$ and $s$ to determine the masses and hence measure $I_\mathrm{A}$ from $\dot{\omega}^\text{\,intr}$ as in \cite{Kehl}. Instead, a self-consistent method is employed to solve for the masses ($m_\mathrm{A}, m_\mathrm{B}$) and  $I_\mathrm{A}$ jointly from $\dot{P}_\mathrm{b}^\text{ intr}(m_\mathrm{A}, m_\mathrm{B}, I_\mathrm{A})$, $s\,(m_\mathrm{A}, m_\mathrm{B})$ and $\dot{\omega}^\text{ intr}\,(m_\mathrm{A}, m_\mathrm{B}, I_\mathrm{A})$. To estimate the probability distribution function for $I_\mathrm{A}$, we perform a Monte Carlo simulation to randomise the observed parameters according to their uncertainties. This process is repeated for the measurements from 2018 to 2030.
\begin{figure}
  \includegraphics[width=\columnwidth]{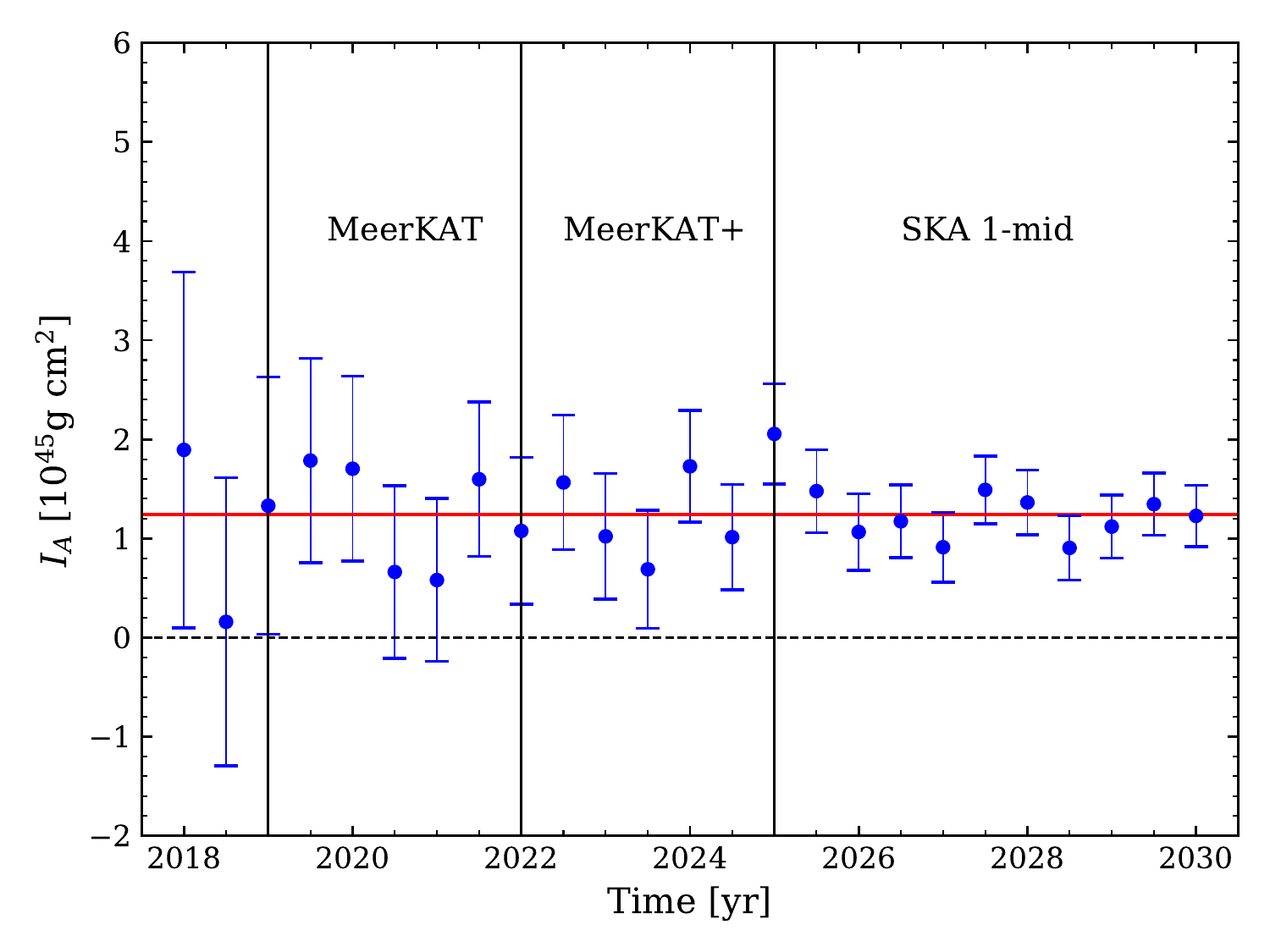}
  \caption{Simulated measurements of the MOI of PSR J0737$-$3039A with time, where two full orbits observation per month are assumed. The red line indicates the theoretical value of the MOI for the chosen EOS AP4 ($I_\mathrm{A}^{\mathrm{AP4}}$).
  \label{fig:IA_yr} }
\end{figure}
\begin{figure}
    \includegraphics[width=\columnwidth]{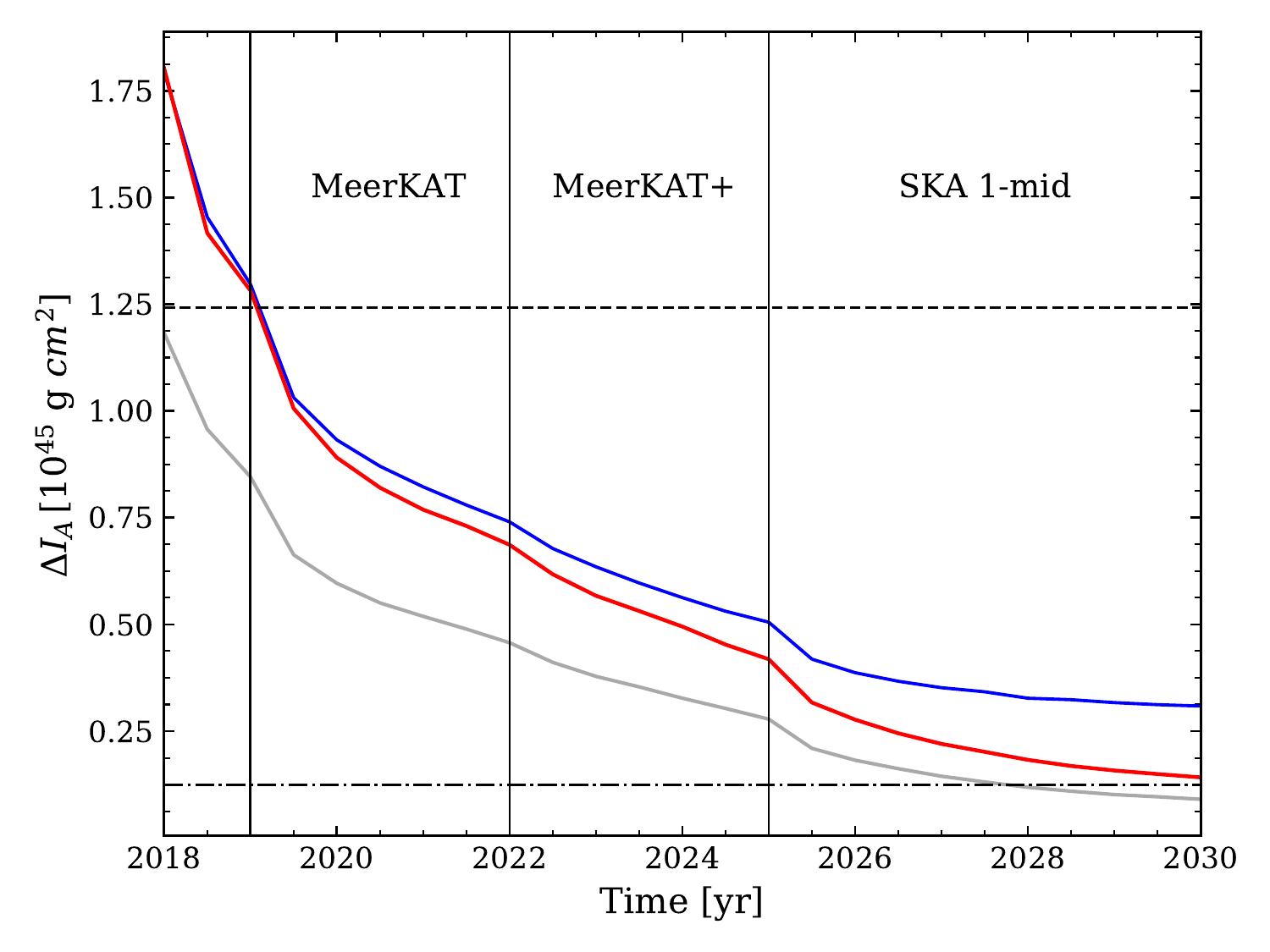}
    \caption{Predicted uncertainty of $I_\mathrm{A}$ as a function of time. The blue line adopts the Galactic measurements ($R_0$, $\Theta_0$) by \protect\cite{Gravity2019} and slope measurement by \protect\cite{Reid14}, whereas the red line assumes no errors in the Galactic model. The grey line is same as the red line but ignores the influence of mass loss to the orbital period change. The theoretical value $I_\mathrm{A}^{\mathrm{AP4}}$ is indicated by the dashed black line, whereas the dash-dotted line is 10\% of the theoretical value. \label{fig:d_IA}}
\end{figure}

Figure~\ref{fig:IA_yr} shows the predicted measurements of $I_\mathrm{A}$ with time, where the new telescopes clearly help to narrow down the uncertainty of $I_\mathrm{A}$. Here we adopt the Galactic measurements ($R_0$, $\Theta_0$) by \cite{Gravity2019} and the slope measurement by \cite{Reid14}. The predicted uncertainty of $I_\mathrm{A}$ with time is also illustrated as the blue line in Figure~\ref{fig:d_IA}. In this case, we expect to achieve an MOI measurement with 25\% precision at 68\% C.L. by the year 2030. Our simulation shows that, although the uncertainty of $\dot{P}_\mathrm{b}^\mathrm{\,obs}$ is initially higher than the Galactic acceleration, it decreases with additional years of precise timing observations (see Figure~\ref{fig:frac_error}), and by 2030, the error in the Galactic acceleration is three times higher than the error in $\dot{P}_\mathrm{b}^\mathrm{\,obs}$, which becomes the limiting factor for measuring the MOI. 

However, the measurements of the Galactic potential is expected to improve through various observational methods, such as Gaia mission \citep{GAIA16} and ongoing observations of Galactic masers \citep{Reid14}. A recent study by \cite{Eilers2019} provides a precise measurement of the circular velocity curve of the Milky Way from 5 to 25 kpc. With the distance from the Sun to the Galactic center $R_0 = 8.122 \pm 0.031$ kpc \citep{Gravity2018}, they determine the rotation speed of the local standard of rest $\Theta_0 = 229.0\pm0.2\,\mathrm{km\,s^{-1}}$, with a slope of $-1.7 \pm 0.1 \,\mathrm{km \, s^{-1} \, kpc^{-1}}$ (statistical errors), corresponding to $b_0 = 0.0603 \pm 0.0035$. The total uncertainties (including systematic errors) given by \cite{Eilers2019} are similar to the measurements used in the previous case (blue line), but here we assume the systematic errors can be well understood in the near future, and only consider the statistical errors. With this assumption, we expect to measure the MOI with 11\% precision at 68\% C.L. in 2030. This is nearly the same as using an error-free Galactic model, which is indicated by the red line in Figure~\ref{fig:d_IA}. Therefore, with future measurements of the Galactic potential and a better understanding of the systematic errors, a MOI measurement with 11\% precision from the Double Pulsar seems realistic.

One important factor for the result is the influence of the mass loss in pulsar A, which was neglected in the previous study by \cite{Kehl}. Without considering this contribution, the uncertainty of $I_\mathrm{A}$ significantly reduces and reaches 7\% by 2030 (see the grey line in Figure~\ref{fig:d_IA}), in contrast to the red line. In addition, we find that increasing the observing cadence does not significantly improve the precision of MOI measurements.

As mentioned in Section~\ref{subsec:Galactic}, different approaches provide very different measurement of the distance of the Double Pulsar, and a compromise distance of 0.8\,kpc is thereby employed in our study. To investigate how distance influences the MOI measurement, we consider two extreme cases, $d=0.4$\,kpc and $d=1.6$\,kpc, with the same setups as in the $d=0.8$\,kpc simulations. Using the current Galactic measurements, we find that the uncertainty of the MOI measurement reaches 17\% by 2030 when $d=0.4$\,kpc, and has a much higher uncertainty (43\%) when $d=1.6$\,kpc. However, with negligible error in the Galactic potential, both predict $\sim$11\% measurements by 2030, same as for the case of $d=0.8$\,kpc. Since an improved Galactic model is expected in the near future, the value we employ for the distance should not have a significant impact on the prediction of the MOI uncertainty.

\begin{figure}
    \includegraphics[width=\columnwidth]{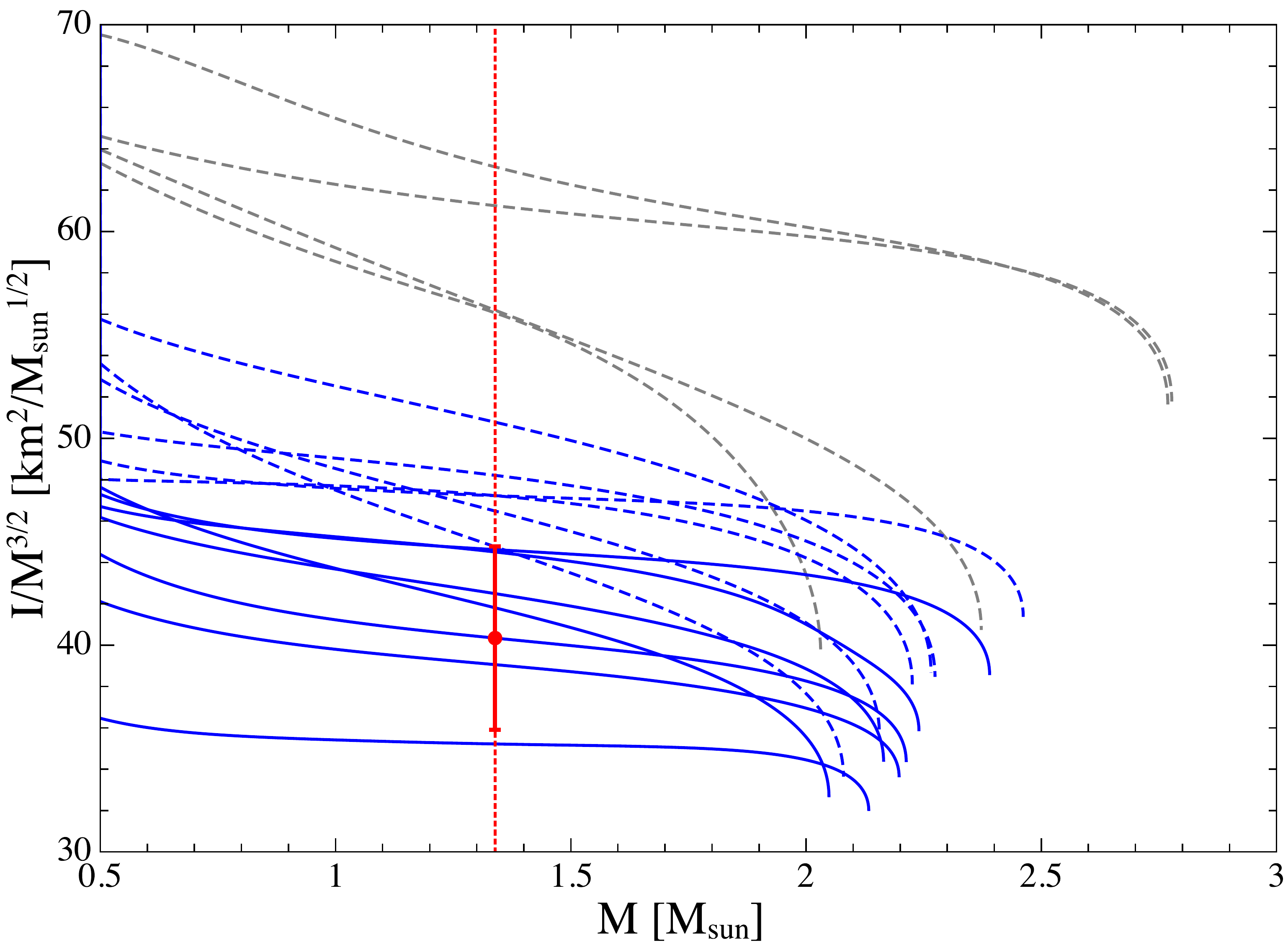}
    \caption{Constraints of EOSs from an 11\% measurement of the MOI of PSR~J0737$-$3039A (red). EOS AP4 was assumed in the simulation (curve through red dot). The grey dashed curves indicate EOSs that are disfavoured by the LIGO/Virgo observations of the GW170817 binary neutron-star merger \citep{LIGO2018}. The blue dashed curves are additionally excluded by the refined (combined with nuclear theory) GW170817 analysis by \citet{Capano2020}. The following EOSs have been plotted (ascending in their intersection with the left border): WFF1, WFF2, AP4, BSk20, AP3, SLy4, BSk25, MPA1, BSk21, SLy9, BL, BSk22, H4, PAL1, MS2, MS0 (\url{https://compose.obspm.fr}). All these EOSs are able to support a NS of $1.98\,M_\odot$, the current lower limit for the maximum mass (see Section~\ref{sec:intro} for details).
    \label{fig:M-I}}
\end{figure}

An 11\% precision measurement of the MOI would further improve the constraints of the EOS of NSs \citep{LS05, Greif_2020}.
Figure~\ref{fig:M-I} shows the MOIs of a number of EOSs, which are scaled by a factor of $M^{3/2}$ in order to reduce the range of the ordinate \citep[cf.][]{LS05}. The $11\%$ measurement predicted from our simulation is illustrated by the red bar centered at the assumed EOS AP4, and located at the precisely measured mass of pulsar A. To compare with the constraints from other methods, we mark the curves in different styles. The observations of the binary neutron-star merger event GW170817 by LIGO/Virgo \citep{LIGO2018} placed a constraint for the radii of both NSs, $11.9\pm1.4\,$km (90\% C.L.), which excludes the EOSs in grey dashed curves. Recently, a more stringent constraint combining GW170817 with nuclear theory was obtained by \cite{Capano2020}, where they found the radius for a $1.4 M_{\odot}$ NS is $11.0_{-0.6}^{+0.9}\,$km (90\% C.L.). This further excludes the EOSs in blue dashed curves. The remaining promising EOSs from this constraint are marked in blue solid curves, which is already very close to our 11\% prediction from the MOI measurement in 2030. With more and more binary NS mergers expected to be detected in the coming years, tighter constraints on the EOS are likely to be achieved. Meanwhile, recent NICER observation delivered a joint mass-radius measurement for PSR J0030$+$0451 from two independent analyses. \cite{Riley2019} found an inferred mass and equatorial radius of $1.34^{+0.15}_{-0.16} M_{\odot}$ and $12.71_{-1.19}^{+1.14}\,$km (68\% C.L.), while \cite{Miller2019} found $1.44^{+0.15}_{-0.14} M_{\odot}$ and $13.02_{-1.06}^{+1.24}\,$km. This is a weak constraint on the EOS, but is expected to improve with more observations in the near future. The upcoming X-ray missions, such as eXTP \citep{eXTP} and ATHENA \citep{athena}, are also promising to improve our understanding of the mass-radius relation for NSs. 

Therefore, it is fair to assume that the GWs and X-ray observations will place a more stringent constraint on the EOS within the next 10 years, and if the EOS can be known with sufficient precision, we can in turn use this information as an input to our analysis, test the LT precession and constrain theories of gravity with the Double Pulsar. We will discuss this scenario in detail in the next section.

\section{Testing Lense-Thirring precession}
\label{sec:LT}

As discussed in the previous section, the MOI measurement of PSR~J0737$-$3039A is expected to reach 11\% accuracy by 2030, whereas GWs and X-ray observations are likely to give a better constraint on the EOS. In this section, we discuss the prospects of testing LT precession and constraining theories of gravity using the Double Pulsar, if the EOS is known.

We again adopt EOS AP4 and this time assume that a precision of 5\% could be achieved when calculating the MOI of pulsar A, based on a (hypothesized) future improvement in our understanding of super-dense matter. Given $I_\mathrm{A}$ as an input to our simulations, only the masses are unknown for the intrinsic orbital period decay $\dot{P}_\mathrm{b}^\text{ intr}$ and the Shapiro shape parameter $s$. With the masses measured from ($\dot{P}_\mathrm{b}^\text{intr}$, $s$) and the given $I_\mathrm{A}$, we can directly test the LT contribution to the periastron advance $\dot{\omega}^\text{LT,A}$. To discuss the physical meaning of such a test, we use the generic framework for relativistic gravity theories introduced by \cite{DT92}, which is fully conservative and based on a Lagrangian that includes a generic term $L_{\rm SO}$ for spin-orbit interaction. As in \cite{DT92}, we will make no assumption about the (strong-field) coupling function $\Gamma_\mathrm{A}^\mathrm{B}$, which enters $L_{\rm SO}$. 
Since the spin axis of pulsar A has been found to be practically parallel to the orbital angular momentum, the general form of the LT contribution to the periastron advance can be written as
\begin{align}
\dot{\omega}^\text{LT,A} = -\frac{2 n_\mathrm{b}^2 I_\mathrm{A} \Omega_\mathrm{A}}{(1-e_\mathrm{T}^2)^{3/2} M} \frac{\sigma_\mathrm{A}}{\mathcal{G}} \,,
\end{align}
where $\sigma_\mathrm{A}$ is a generic strong-field spin-orbit coupling constant, defined by
\begin{align}
\sigma_\mathrm{A} = \frac{1}{c^2} \left[\Gamma_\mathrm{A}^\mathrm{B} +\left(\Gamma_\mathrm{A}^\mathrm{B}-\frac{1}{2}\mathcal{G}\right)\frac{m_\mathrm{B}}{m_\mathrm{A}}\right] \,.
\label{eq:sigma_A}
\end{align}
In GR, the generalised gravitational constant $\mathcal{G}$ equals $G$, and the coupling function $\Gamma_\mathrm{A}^\mathrm{B}$ equals $2G$ \citep{DT92}, so that
\begin{align}
\sigma_\mathrm{A}^\mathrm{GR} = \frac{G}{c^2} \left(2+ \frac{3}{2}\frac{m_\mathrm{B}}{m_\mathrm{A}}\right) \,.
\label{eq:sigma_A^GR}
\end{align}
But in other theories, $\Gamma_\mathrm{A}^\mathrm{B}$ is expected to deviate from $2G$, including modifications by self-gravity contributions from the strongly self-gravitating masses in the system.

We define a parameter $\delta_\mathrm{LT}$ to measure the relative deviation of the theory-independent parameter $\sigma_\mathrm{A}/\mathcal{G}$ from its GR prediction,
\begin{align}
\delta_\mathrm{LT} = 
\left(\frac{\sigma_\mathrm{A}}{\mathcal{G}}\right) 
\left(\frac{\sigma_\mathrm{A}^\mathrm{GR}}{G}\right)^{-1} - 1\,.
\end{align}
By inserting Eq.~\eqref{eq:sigma_A} into the above definition, one obtains for the spin-orbit coupling function
\begin{align}
\frac{\Gamma_\mathrm{A}^\mathrm{B}}{2\mathcal{G}} -1 = \left(\frac{3 + x_\mathrm{A}}{4}\right) \delta_\mathrm{LT} \,,
\end{align}
To assess potential constraints on a non-GR spin-orbit coupling, we multiply the expression of $\dot{\omega}^\text{LT,A}$ in GR (last term in Eq.~\eqref{eq:omdot}) by $(1+\delta_\mathrm{LT})$, and solve for the parameter $\delta_\mathrm{LT}$ using the three PK parameters $\dot{P}_\mathrm{b}^\text{ intr}(m_\mathrm{A}, m_\mathrm{B})$, $s\,(m_\mathrm{A}, m_\mathrm{B})$, and $\dot{\omega}^\text{intr}(m_\mathrm{A}, m_\mathrm{B}, \delta_\mathrm{LT})$. One has to keep in mind that, for simplicity, we make here the assumption that the non-spin related parts of the orbital dynamics and signal propagation are (to sufficient approximation) given by their GR expressions. It goes without saying, that in practice one has to conduct a fully self-consistent analysis within a given class of alternative gravity theories. For a discussion that purely focuses on the measurability of a potential deviation in the LT contribution, our approach is sufficient.

\begin{figure}
    \includegraphics[width=\columnwidth]{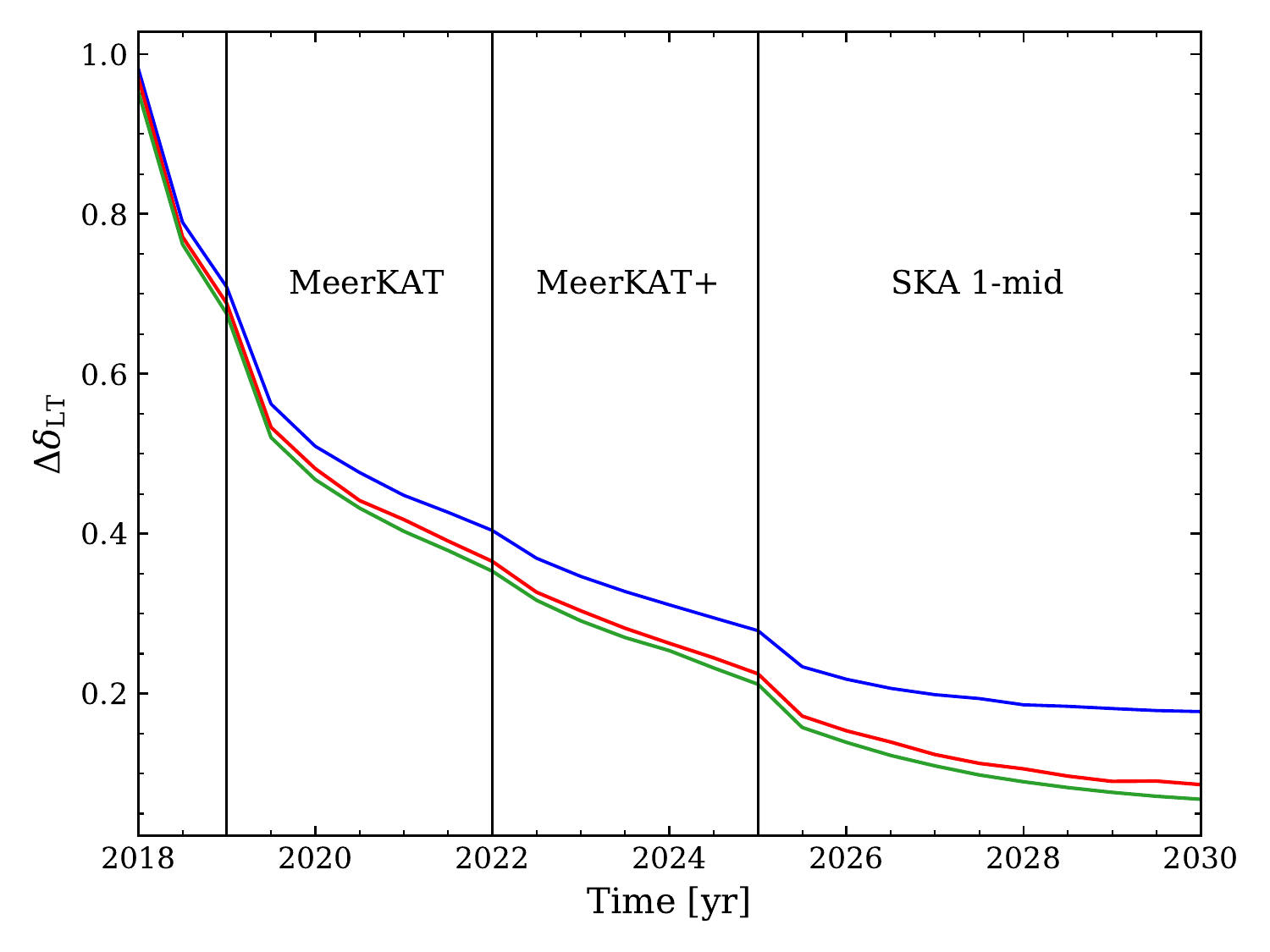}
    \caption{Predicted uncertainty of $\delta_\mathrm{LT}$ as a function of time. The blue line adopts Galactic measurements ($R_0, \Theta_0$) by \citet{Gravity2019} and slope measurement by \citet{Reid14}, the red line adopts Galactic and slope measurements by \citet{Eilers2019}, and the green line assumes no errors in the Galactic model, the distance and the MOI.
    }
    \label{fig:d_lambda}
\end{figure}

Figure~\ref{fig:d_lambda} shows the expected decrease in the uncertainty of $\delta_\mathrm{LT}$ with future observations. With $R_0$ and $\Theta_0$ measurements from \cite{Gravity2019} and the slope measurement from \cite{Reid14}, we expect to measure $\delta_\mathrm{LT}$ with $18\%$ precision at 68\% C.L. by 2030, which is indicated by the blue line. The red line adopts the Galactic measurements from \cite{Eilers2019}, where we expect to achieve a 9\% precision by 2030. In the ideal case, we assume that the Galactic potential, the distance to the Double Pulsar, and the MOI can be precisely measured in the future, so that we could leave out the errors. In this scenario, we expect to measure $\delta_\mathrm{LT}$ with 7\% precision by 2030 (green line). We have seen in Section~\ref{sec:MOI} that change from the Galactic measurements by \cite{Eilers2019} to an error-free Galactic model has little enhancement on the measurements of the MOI, and the uncertainty of the timing parallax is relatively small, therefore, the improvement from 9\% (red line) to 7\% (green line) is to a fair fraction (nearly half) related to the uncertainty of the MOI.

\cite{Breton2008} have conducted a different experiment for spin-orbit coupling in the Double Pulsar system. Studying the geodetic precession of pulsar B, they were able to show that $\sigma_\mathrm{B}/\mathcal{G}$ is in agreement with GR, with a precision of about 13\%. Analogously to Eq.~(\ref{eq:sigma_A}), $\sigma_\mathrm{B}$ is related to $\Gamma_\mathrm{B}^\mathrm{A}$. A priori there is no reason to assume that $\Gamma_\mathrm{B}^\mathrm{A}$ and $\Gamma_\mathrm{A}^\mathrm{B}$ are equal \citep[see discussion in][]{DT92}. Consequently, a LT test with pulsar A would nicely complement the geodetic precession test of \cite{Breton2008}, when investigating the relativistic interaction between the proper rotation of the two NSs and their orbital motion.

Finally, short range modifications of gravity, related to the strong gravitational field of a NS, could significantly change the structure of the star and therefore its MOI, without any ``direct'' impact on the orbital dynamics or the signal propagation in a binary pulsar system. Examples of such theories are scalar-tensor theories with a massive scalar field having a sufficiently short Compton wavelength \citep[see e.g.][]{RP16,YDP16}. While in such a scenario, PK parameters related to time dilation, GW damping, and Shapiro delay remain (practically) unaffected \citep[see e.g.][]{Alsing_2012}, one could still expect a deviation in the precession of periastron of the Double Pulsar. The reason is that due to the modification of the MOI the spin of pulsar A and therefore the spin-orbit coupling is modified. Testing the LT precession in the Double Pulsar can therefore be used to constrain such deviations from GR. It is important to note, that $\dot{P}_\mathrm{b}^{\dot{m}}$ would also be modified accordingly, and therefore has to be accounted for. Hence, limits on $\delta_\mathrm{LT}$ would consequently be somewhat weaker than given above (cf.\ Section~\ref{sec:MOI}). In such a scenario it could generally be difficult to disentangle uncertainties in the EOS and deviations from GR by astronomical observations. For this, a combination of various experiments, like GWs from binary neutron-star mergers, X-ray observations, and radio pulsar timing might turn out to be necessary. Nonetheless, the future measurement of the LT precession in the Double Pulsar is expected to provide important contributions when constraining such deviations from GR.


\section{Next-to-leading order gravitational wave damping}
\label{sec:3p5}

In GR, the loss of energy of a material system due to GWs is to leading order sourced by a time-dependent mass quadrupole \citep{Einstein:1918btx,Eddington:1922ds}. This also holds for binary systems where a change in the mass quadrupole is driven by gravity itself. It enters the two-body equations of motion at the 2.5PN order \citep[see e.g.][]{Damour_1987}. When computing the next-to-leading order contribution to GW damping, one also has to account for the mass-octupole and the current quadrupole moments \citep{Thorne:1980ru}. Next-to-leading order contributions enter the equations of motion at 3.5PN ($\mathcal{O}(c^{-7})$), and therefore correspond to the 1PN corrections in the radiation reaction force \citep{Iyer_1995,Pati_2002, Konigsdorffer_2003,Nissanke_2005}. The corresponding change in the orbital period of a binary system has been determined out by \cite{BS89} and is given by Eq.~(\ref{eq:pbdot_3p5PN}). In this section we will investigate if next-to-leading order corrections to the GW damping are expected to become important in the near future for the timing observation of the Double Pulsar.

Again we assume EOS AP4 and a 5\% error in the knowledge of the MOI $I_\mathrm{A}$. We implement the 3.5PN contribution into our model by using Eq.~\eqref{eq:pbdot_3p5PN}, and adjust the TOAs accordingly. After running simulations as described in Section~\ref{sec:sim}, we obtain the measured PK parameters. We use Eq.~\eqref{eq:pbdot_GW} to solve for the relative correction of the 3.5PN order $X_\mathrm{3.5PN}$ using the three PK parameters $\dot{P}_\mathrm{b}^\text{ intr}(m_\mathrm{A}, m_\mathrm{B}, X_\mathrm{3.5PN})$, $s\,(m_\mathrm{A}, m_\mathrm{B})$, and $\dot{\omega}^\text{intr}(m_\mathrm{A}, m_\mathrm{B})$.

\begin{figure}
    \includegraphics[width=\columnwidth]{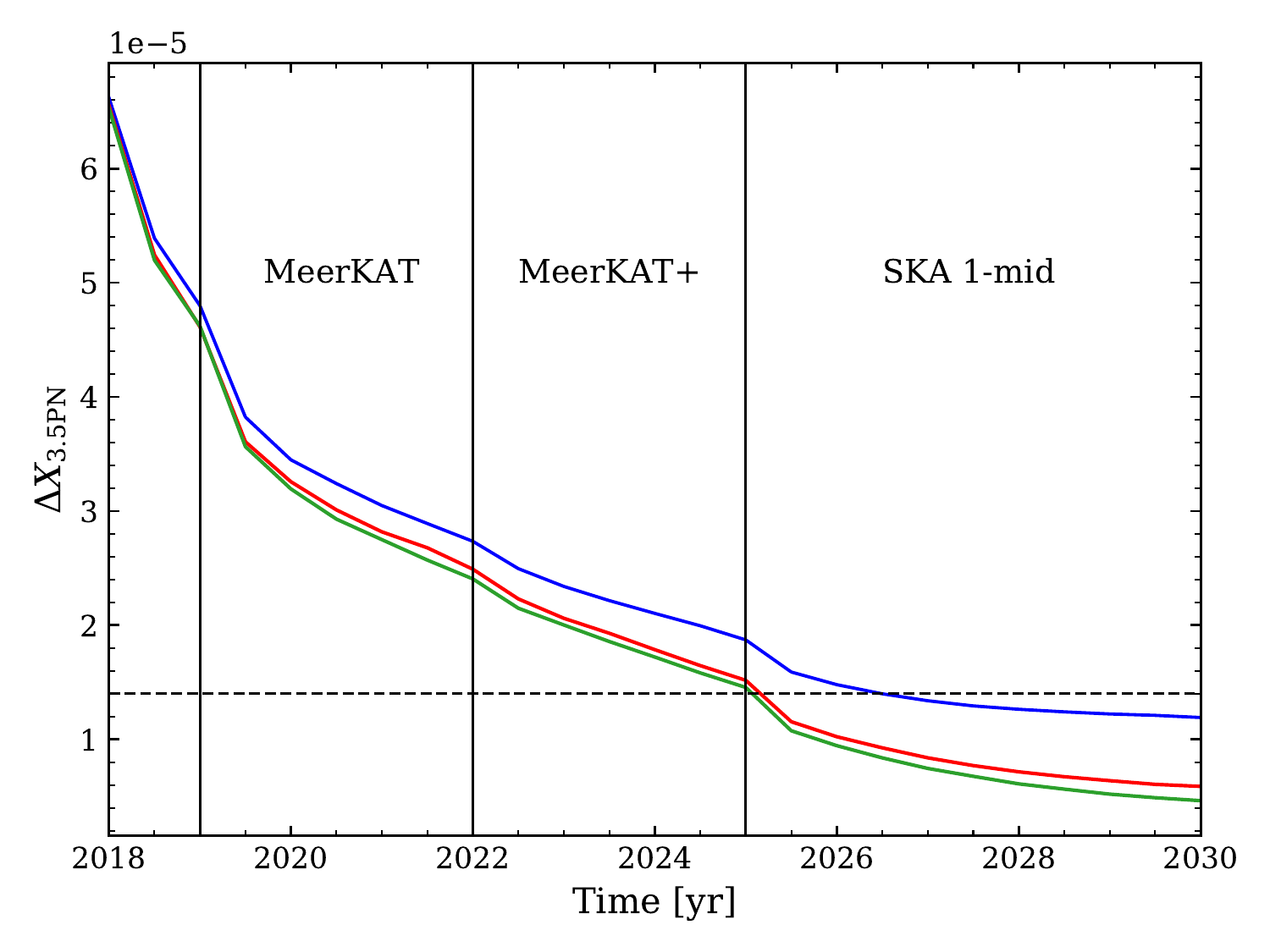}
    \caption{Same as Figure~\ref{fig:d_lambda} but for the uncertainty of the 3.5PN order GW correction $X_{3.5\mathrm{PN}}$. The dashed line denotes the theoretical value of $X_{3.5\mathrm{PN}}$.}
    \label{fig:3p5PN}
\end{figure}

Figure~\ref{fig:3p5PN} illustrates the predicted uncertainty of $X_{3.5\mathrm{PN}}$ with observing phase, which will fall below its theoretical value $X_{3.5\mathrm{PN}}^\mathrm{theo}$ in the SKA1-mid era. The colours of the lines represent the same conditions as in Figure~\ref{fig:d_lambda}. The blue line shows the improvements in $\Delta X_{3.5\mathrm{PN}}$ with Galactic parameters from \cite{Gravity2019} and the slope measurement by \cite{Reid14}, which will reach a precision of 85\% at 68\% C.L. by 2030. Adopting the Galactic measurements (statistical errors) by \cite{Eilers2019}, the red line shows that $X_{3.5\mathrm{PN}}$ can be constrained with a precision of 42\% by 2030. By contrast, in the ideal case where there are no errors in the Galactic model, the distance and the MOI, $X_{3.5\mathrm{PN}}$ can be constrained with a precision of 33\% by 2030, where nearly half of the improvement is contributed from the MOI.

\section{Potential new discoveries}
\label{sec:new}

Large pulsar surveys with MeerKAT, FAST and the forthcoming SKA, such as TRAPUM \citep{SK2016} and CRAFTS \citep{Li2018}, can potentially discover more relativistic double neutron star (DNS) systems, preferably with a more compact orbit than PSR~J0737$-$3039. An example of such a system, PSR~J1946$+$2052, with a more relativistic orbit than the Double Pulsar ($P_\text{b} \simeq 1.88$~h) and larger periastron advance ($\dot\omega \approx 26\,{\rm deg\,yr^{-1}}$) and LT precession ($\dot\omega_{\rm LT} \approx 0.001\,{\rm deg\,yr^{-1}}$), was recently discovered in the PALFA survey \citep{Stovall2018}. In its orbital parameters, the PSR~J1946$+$2052 system resembles a system similar to the Double Pulsar, but that has evolved further due to GW damping, by about 40\,Myr. While it is still unclear, if for PSR~J1946$+$2052 the necessary precision in the mass determination can be reached to rival the Double Pulsar in the tests proposed here \footnote{Since PSR~J1946$+$2052 is less luminous compared to the Double Pulsar, and $s$ is not measurable due to its orientation.}, it certainly adds confidence to the hope of finding more relativistic ``cousins'' of the Double Pulsar in the coming years. Such binary pulsars would quite likely enable MOI measurements with superior precision within a comparably short period of time, and improve the constraints of the EOS.

Here we consider two scenarios, one with an orbital period of 100 minutes and one with 50 minutes, which are within the expected acceleration searches by MeerKAT. Assuming such systems can be found in 2020 and we start timing them regularly from 2021, with two orbits per month, we run our simulation again to predict the measurements of the MOI. To simplify the simulation, we assume these systems satisfy the conditions of the Double Pulsar (inclination $i$ close to 90 degrees, similar distance and brightness, etc.) but with modified orbital parameters, assuming that these systems had an orbit like the Double Pulsar some time in the past, and then evolved by GW damping to an orbital period of 100 or 50 minutes. In reality, these systems are likely to be further away. Nonetheless, it is also possible that such systems are bright and nearby, but were missed in the past surveys due to their high acceleration \citep[see][]{johnston91,ransomphd,jouteux,Ng2014, Cameron2018}. 

We calculate the evolved semi-major axis using Kepler's third law and the evolved eccentricity using the $a-e$ relation in \cite{Peters1964}, for the orbital period of 100 minutes and 50 minutes, respectively. Then we calculate the PK parameters and run simulations as described in Section \ref{sec:sim} and \ref{sec:MOI}. Assuming the same distance as the Double Pulsar, we convert the uncertainty of timing parallax into an uncertainty for the distance. The Galactic measurement by \cite{Eilers2019} is adopted in the simulation and, as before, we assume the systematic uncertainties can be well understood in the future.

Our results show that, for the DNS system with an orbital period of 100 minutes, we could measure the MOI with 12\% precision by 2030 and with 4.5\% by 2035 at 68\% C.L. As for an orbital period of 50 minutes, we expect an MOI measurement with 1.5\% precision by 2030 and with 0.5\% by 2035 at 68\% C.L. Such measurements would probably be comparable to the by then available constraints from other methods (GWs and X-ray observations, nuclear physics, etc.) and help for determining the EOS of NSs.

Furthermore, LISA has the potential to discover ultra relativistic DNS systems with a characteristic orbital frequency of 0.8 mHz \citep{Lau20}. \cite{Thrane2020} suggested that following up such systems with SKA for 10 years could potentially measure the mass-radius relation with a precision <1\%. To this end, we perform a simulation for a DNS system with 20 minute orbital period, and find an MOI precision of $\sim$0.2\% (68\% C.L.) may be possible with 10 years of timing with SKA 1-mid.

However, there is a low chance that the new discovered DNS systems will be edge-on to our line-of-sight, as is the case for PSR~J0737$-$3039, hence a precise measurement of $s$ might not be possible. Instead, we may need to use $\gamma$ to constrain the masses and MOI, whose fractional error is usually a few orders of magnitude larger than $s$ (see Figure~\ref{fig:frac_error}). This is indeed the case for PSR~J1946$+$2052, despite its relativistic nature, determining the masses with sufficient precision will be challenging.

Moreover, not all DNS systems are ideal to test the Lense–Thirring precession in terms of periastron advance $\dot{\omega}^\mathrm{LT}$. Systems like the aforementioned PSR~J1757$-$1854 have a large eccentricity most likely caused by a large kick \citep{tauris17} causing a significant misalignment between the spin of pulsar and the orbital angular momentum, and hence $\dot{\omega}^\mathrm{LT}$ can not be determined as straightforwardly as in the Double Pulsar. However, as pointed out in Section \ref{sec:methodSO}, this allows an alternate test using the contribution of LT precession to the rate of change of the projected semi-major axis $\dot{x}^\mathrm{LT}$ \citep{Cameron2018} if profile changes due to geodetic precession can be accounted for in the timing process and the spin orientation can be determined with sufficient precision.


\section{Conclusion}
\label{sec:dis}

In this paper, we have developed a consistent method to measure the MOI of radio pulsars, which has been applied to mock data for the Double Pulsar. We simulated TOAs of PSR~J0737$-$3039A assuming future observations with MeerKAT, MeerKAT+ and SKA 1-mid which cover two orbits per month. We found a MOI measurement with 11\% accuracy (68\% C.L.) could be achievable by the end of this decade, if we have sufficient knowledge of the Galactic gravitational potential (e.g., from Gaia mission \citep{GAIA16}). We also found that the mass loss of pulsar A has a considerable impact on the measurement of the MOI. Neglecting this contribution to the orbital period change leads to an overoptimistic prediction. This is the main reason why, even with the better timing precision used in this paper as compared to \cite{KW2009}, by $\sim$2030 we would still only reach the same accuracy level as predicted by \cite{KW2009}. Additionally, the assumptions made in this paper are more realistic compared to \cite{Kehl}, with timing precision from MeerKAT observation, as well as updated timeline and size of upcoming telescopes.

In the second part of the paper, Section~\ref{sec:LT} and \ref{sec:3p5}, we have assumed that a better constraint on the EOS might be achieved with GWs and X-ray observations in the future, so as to investigate the capability of testing LT precession and 3.5PN order contributions to the GW damping. This assumption coincides with \cite{Landry2020} where they found that constraints from GWs and X-ray observations are likely to have larger contributions in constraining the EOS than the MOI measurement of J0737$-$3039A. Assuming a 5\% error in the determination of the MOI, we simulated measurements of the relative deviation of the theory-independent spin-orbit coupling parameter $\sigma_\mathrm{A}/\mathcal{G}$ from GR's prediction. We found a 9\% precision measurement is possible by 2030 with an improved Galactic model, whereas a 7\% precision measurement in the ideal case --- no errors in the Galactic model, the distance, and the MOI. This test is a complement to \cite{Breton2008}, where they found a 13\% constraint on $\sigma_\mathrm{B}/\mathcal{G}$. This measurement would enable a constraint for the coupling function $\Gamma_\mathrm{A}^\mathrm{B}$ that enters the spin-orbit Lagrangian of the two-body interaction for strongly self-gravitating masses. Such a measurement could be sensitive to short range deviations from GR, which otherwise would not show up in the orbital dynamics of such systems.

We have also studied the measurability of GR's next-to-leading (3.5PN) order GW-damping contribution. We predicted that the uncertainty of the 3.5PN order correction $X_{3.5\mathrm{PN}}$ will fall below its theoretical value at the beginning of SKA 1-mid ($\sim$2026) and a measurement of $X_{3.5\mathrm{PN}}$ with $3\sigma$-significance is possible in $\sim$10 years, if by then we have sufficient knowledge of the Galactic gravitational potential, pulsar distance, and the EOS. This means that from the SKA 1-mid era, we will have to include the 3.5PN term in our analysis in order to avoid any bias. Binary mergers detected by LIGO/Virgo do provide constraints on post-Newtonian (PN) terms \citep{LIGO2016GR}. Their way of counting the PN contributions is relative to the Einstein quadrupole formula, i.e. the order they enter the radiation reaction force \citep{Blanchet_LRR}. Their 1PN term therefore contains 3.5PN contributions from the equations of motion. As a comparison to our 3.5PN 3-$\sigma$ result, \citep{LIGO2019GR} provide a $\sim$10\% measurement (90\% C.L.) of the (radiative) 1PN coefficient with GW170817. Future merger events will most likely lead to even more precise measurements of this term. While at the 2.5PN (0PN radiative) level, the Double Pulsar is still many orders of magnitude more precise than LIGO/Virgo mergers \citep[][Kramer et al. in prep.]{Kramer_2016}. When it comes to higher order PN contributions, we conclude that binary pulsars are not expected to be competitive, simply because of the much smaller orbital velocity.

Finally, we discussed potential new discoveries of DNS systems with radio telescopes like MeerKAT, FAST, and SKA, as well as the space-based future GW observatory LISA. We demonstrated that for a DNS system which mimics the evolved PSR~J0737$-$3039 with an orbital period of 50 minutes, the MOI measurement is expected to reach 1.5\% precision (68\% C.L.) after 10 years observation with MeerKAT, MeerKAT+ and SKA 1-mid, and 0.5\% precision after 15 years. Moreover, LISA is expected to find DNS systems with a characteristic orbital period of 20 minutes in the near future \citep{Lau20}. Such discoveries can significantly tighten the constraints for the EOS.

To conclude, although the EOS constraints resulting from a future MOI measurement with the Double Pulsar are not likely to exceed those with LIGO/Virgo mergers and X-ray observations in the coming years, we still anticipating other aspects of science coming from this unique gravity laboratory in future studies based on an improved understanding of the NS EOS as an input. Furthermore, the discovery of more relativistic binary pulsars, possible with the unprecedented surveying capabilities of new and upcoming radio telescopes and advances in data analysis \citep[e.g.][]{cobra}, could ultimately lead to EOS constraints quite competitive with other methods.

\section*{Acknowledgements}

We are grateful to Vivek Venkatraman Krishnan and Aditya Parthasarathy for helpful discussions, and Paulo Freire for carefully reading the manuscript. We would also like to thank our reviewer, Gerhard Sch\"afer, for his helpful comments. HH is a member of the International Max Planck Research School for Astronomy and Astrophysics at the Universities of Bonn and Cologne. NW and MK gratefully acknowledge support from ERC Synergy Grant “BlackHoleCam” Grant Agreement Number 610058. This work has been supported by the Chinese Academy of Sciences (CAS) and the Max-Planck Society (MPG) in the framework of LEGACY cooperation on low-frequency gravitational wave astronomy.

\section*{Data availability}

The data underlying this article will be shared on reasonable request to the corresponding author.


\bibliographystyle{mnras}
\bibliography{main} 




\bsp	
\label{lastpage}
\end{document}